\documentclass[aps,prd,preprint,superscriptaddress,showpacs,floatfix,11pt]{revtex4}

\usepackage[dvips]{graphics}
\usepackage[dvips]{graphicx}
\usepackage{psfrag}
\usepackage{longtable}
\usepackage{amsfonts}
\usepackage{mathrsfs}

\begin{document}

\preprint{BNL-HET-04/14, COLO-HEP-502}
\preprint{August, 2004}
\vskip 1.5in

\title{Lepton Flavor Violating Decays, Soft Leptogenesis and SUSY SO(10)}

\author{Mu-Chun Chen}
\email[]{chen@quark.phy.bnl.gov}
\affiliation{High Energy Theory Group, Department of Physics, 
Brookhaven National Laboratory, Upton, NY 11973, U.S.A.}
\author{K.T. Mahanthappa}
\email[]{ktm@pizero.colorado.edu}
%\thanks{}
\affiliation{Department of Physics, University of Colorado, 
Boulder, CO80309-0390, U.S.A.}


%\date{\today}

\begin{abstract}
We investigate lepton flavor violating decays in a SUSY SO(10) model 
with symmetric 
textures recently constructed by us. Unlike the models with lop-sided 
textures which 
give rise to a large decay rate for $\mu \rightarrow e \gamma$, the decay rate 
we get is much suppressed and yet it is large enough to be accessible 
to the next generation 
of experiments. We have also investigated the 
possibility of baryogenesis resulting from 
soft leptogenesis. We find that 
with the soft SUSY masses assuming their natural values, $B^\prime 
\equiv \sqrt{BM_{1}} \sim 1.4 \; TeV$
and $Im(A) \sim 1 \; TeV$, 
the observed baryon asymmetry in the Universe can be accommodated in our model.
We have also updated the predictions of our model 
for the masses, mixing angles and CP violating 
measures in both charged fermion and neutrino sectors, using the most up-to-date 
experimental data as input.

\end{abstract}


%\keywords{}

%\maketitle must follow title, authors, abstract and \pacs
\maketitle

%{\bf Introduction}
%\label{}
%\subsection{}
%\subsubsection{}

\section{Introduction}

After Neutrino 2004, the allowed region for the neutrino oscillation parameters 
has been reduced significantly, and their measurements have now entered the precision phase.
There have been a few supersymmetric (SUSY) 
$SO(10)$ models constructed aiming to accommodate the 
observed neutrino masses and mixing angles 
(For a recent review on SO(10) models, see Ref.~\cite{Chen:2003zv}.) 
By far, the LMA solution is the most difficult to obtain. 
Most of the models in the literature assume ``lopsided'' mass matrices. 
In our model based on SUSY $SO(10) \times SU(2)$~\cite{Chen:2000fp}%,Chen:2001pr,Chen:2002pa} 
(referred to ``CM'' herein), we consider {\it symmetric} mass matrices which  
result from the left-right symmetric breaking of $SO(10)$ and the breaking of 
family symmetry $SU(2)$. In view of the much improved experimental data on  
neutrino oscillation parameters as well as those in the quark mixing from B Physics, 
we re-analyze our model and find that it can still accommodate 
all experimental data within $1 \sigma$. We investigate several lepton flavor 
violating (LFV) processes in our model, including the decay of the muon 
into an electron and a photon, which is the most stringently 
constrained LFV process. We also investigate in this paper  
the possibility of baryogenesis utilizing 
soft leptogenesis. 

This paper is organized as follows: 
In Sec. \ref{model}, we briefly describe our model, and show 
its predictions for the masses, mixing angles and CP violating phases in both 
charged fermion and neutrino sectors, using the most up-to-date 
experimental data as input. 
Various decay rates for lepton flavor violation processes  are 
calculated in Sec. \ref{lfv}. 
Sec. \ref{leptogenesis} concerns soft leptogenesis 
in our model, while Sec. \ref{concl} concludes this paper.

\section{The Model}\label{model}

The details of our model based on $SO(10) \times SU(2)_{F}$ are contained 
in CM~\cite{Chen:2000fp}.%,Chen:2001pr,Chen:2002pa}.  
The following is 
an outline of its salient features. In order to
specify the superpotential uniquely, we invoke 
$Z_{2} \times Z_{2} \times Z_{2}$ discrete symmetry. The matter fields are
\begin{displaymath} 
\psi_{a} \sim (16,2)^{-++} \quad (a=1,2), \qquad 
\psi_{3} \sim (16,1)^{+++} 
\end{displaymath}
where $a=1,2$ and the subscripts refer to family indices; the superscripts 
$+/-$ refer to $(Z_{2})^{3}$ charges. The Higgs fields which break $SO(10)$
and give rise to mass matrices upon acquiring VEV's are
\begin{eqnarray}
(10,1):\quad & T_{1}^{+++}, \quad T_{2}^{-+-},\quad
T_{3}^{--+}, \quad T_{4}^{---}, \quad T_{5}^{+--} \nonumber\\ 
(\overline{126},1):\quad & \overline{C}^{---}, \quad \overline{C}_{1}^{+++},
\quad \overline{C}_{2}^{++-}  
\nonumber
\end{eqnarray}
Higgs representations $10$ and $\overline{126}$ give rise to Yukawa couplings
to the matter fields which are symmetric under the interchange of family
indices. $SO(10)$ is broken through the left-right symmetry breaking chain, and 
symmetric mass matrices arise.
The $SU(2)$ family symmetry~\cite{Barbieri:1997ww} is broken in two steps and
the mass hierarchy is produced using the Froggatt-Nielsen 
mechanism:
%\begin{equation}
%\label{eq:steps} 
$SU(2) \stackrel{\epsilon M}{\longrightarrow} 
U(1) \stackrel{\epsilon' M}{\longrightarrow}
nothing$
%\end{equation}
where $M$ is the UV-cutoff of the effective theory above which the family
symmetry is exact, and $\epsilon M$ and $\epsilon^{'} M$ are the VEV's
accompanying the flavon fields given by
\begin{eqnarray}
(1,2): \quad & \phi_{(1)}^{++-}, \quad \phi_{(2)}^{+-+}, \quad \Phi^{-+-}
\nonumber\\ 
(1,3): \quad & S_{(1)}^{+--}, \quad S_{(2)}^{---}, \quad
\Sigma^{++-} 
\end{eqnarray}
The various aspects of VEV's of Higgs and flavon fields are given in CM.

The superpotential of our model is
\begin{equation}
W = W_{Dirac} + W_{\nu_{RR}}
\end{equation}
\begin{eqnarray}
W_{Dirac}=\psi_{3}\psi_{3} T_{1}
 + \frac{1}{M} \psi_{3} \psi_{a}
\left(T_{2}\phi_{(1)}+T_{3}\phi_{(2)}\right)
\nonumber\\
+ \frac{1}{M} \psi_{a} \psi_{b} \left(T_{4} + \overline{C}\right) S_{(2)}
+ \frac{1}{M} \psi_{a} \psi_{b} T_{5} S_{(1)}
\nonumber\\
W_{\nu_{RR}}=\psi_{3} \psi_{3} \overline{C}_{1} 
+ \frac{1}{M} \psi_{3} \psi_{a} \Phi \overline{C}_{2}
+ \frac{1}{M} \psi_{a} \psi_{b} \Sigma \overline{C}_{2} \; .
\end{eqnarray}
The mass matrices then can be read from the superpotential to be
\begin{eqnarray}
M_{u,\nu_{LR}} & = &
\left( \begin{array}{ccc}
0 & 0 & \left<10_{2}^{+} \right> \epsilon'\\
0 & \left<10_{4}^{+} \right> \epsilon & \left<10_{3}^{+} \right> \epsilon \\
\left<10_{2}^{+} \right> \epsilon' & \left<10_{3}^{+} \right> \epsilon &
\left<10_{1}^{+} \right>
\end{array} \right)
\nonumber\\
 & = & 
\left( \begin{array}{ccc}
0 & 0 & r_{2} \epsilon'\\
0 & r_{4} \epsilon & \epsilon \\
r_{2} \epsilon' & \epsilon & 1
\end{array} \right) M_{U}
\end{eqnarray}
\begin{eqnarray}
M_{d,e} & = & 
\left(\begin{array}{ccc}
0 & \left<10_{5}^{-} \right> \epsilon' & 0 \\
\left<10_{5}^{-} \right> \epsilon' &  (1,-3)\left<\overline{126}^{-} \right>
\epsilon & 0\\ 0 & 0 & \left<10_{1}^{-} \right>
\end{array} \right)
\nonumber\\
 & = & 
\left(\begin{array}{ccc}
0 & \epsilon' & 0 \\
\epsilon' &  (1,-3) p \epsilon & 0\\
0 & 0 & 1
\end{array} \right) M_{D} \; ,
\end{eqnarray}
where
$M_{U} \equiv \left<10_{1}^{+} \right>$, 
$M_{D} \equiv \left<10_{1}^{-} \right>$, 
$r_{2} \equiv \left<10_{2}^{+} \right> / \left<10_{1}^{+} \right>$, 
$r_{4} \equiv \left<10_{4}^{+} \right> / \left<10_{1}^{+} \right>$ and
$p \equiv \left<\overline{126}^{-}\right> / \left<10_{1}^{-} \right>$.
The right-handed neutrino mass matrix is  
\begin{eqnarray}
M_{\nu_{RR}} & = &  
\left( \begin{array}{ccc}
0 & 0 & \left<\overline{126}_{2}^{'0} \right> \delta_{1}\\
0 & \left<\overline{126}_{2}^{'0} \right> \delta_{2} 
& \left<\overline{126}_{2}^{'0} \right> \delta_{3} \\ 
\left<\overline{126}_{2}^{'0} \right> \delta_{1}
& \left<\overline{126}_{2}^{'0} \right> \delta_{3} &
\left<\overline{126}_{1}^{'0} \right> \end{array} \right)
\nonumber\\
 & = & 
\left( \begin{array}{ccc}
0 & 0 & \delta_{1}\\
0 & \delta_{2} & \delta_{3} \\ 
\delta_{1} & \delta_{3} & 1
\end{array} \right) M_{R}
\label{Mrr}
\end{eqnarray}
with $M_{R} \equiv \left<\overline{126}^{'0}_{1}\right>$.
Here the superscripts $+/-/0$ refer to the sign of the hypercharge. 
It is to be noted that there is a factor of $-3$ difference between the $(22)$
elements of mass matrices $M_{d}$ and $M_{e}$. This is due to the CG
coefficients associated with $\overline{126}$; as a consequence, we obtain the
phenomenologically viable Georgi-Jarlskog relation.
We then parameterize the Yukawa matrices as follows, after removing 
all the non-physical phases by rephasing various matter fields: 
\begin{eqnarray}
Y_{u, \nu_{LR}} & = & \left(
\begin{array}{ccc}
0 & 0 & a\\
0 & b e^{i\theta} & c\\
a & c & 1
\end{array}
\right) d
\\
Y_{d,e} & = & \left(
\begin{array}{ccc}
0 & e e^{-i\xi} & 0 \\
e e^{i\xi} & (1,-3) f & 0 \\
0 & 0 & 1
\end{array}
\right) h \quad .
\label{phaseremoved}
\end{eqnarray}

We use the following as inputs at 
$M_{Z}=91.187 \; GeV$~\cite{Eidelman:2004wy,Fusaoka:1998vc}: %,Hocker:2001xe}:
\begin{eqnarray}
m_{u} & = & 2.21 \; MeV (2.33^{+0.42}_{-0.45})\nonumber\\ 
m_{c} & = & 682 \; MeV (677^{+56}_{-61})\nonumber\\
m_{t} & = & 181 \; GeV (181^{+}_{-}13) \nonumber\\
m_{e} & = & 0.486 \; MeV (0.486847)\nonumber\\
m_{\mu} & = & 103 \; MeV (102.75)\nonumber\\
m_{\tau} & = & 1.74 \; GeV (1.7467) \nonumber\\
\vert V_{us} \vert & = & 0.225 (0.221-0.227)\nonumber\\
\vert V_{ub} \vert & = & 0.00368 (0.0029-0.0045)\nonumber\\
\vert V_{cb} \vert & = & 0.0392 (0.039-0.044)\nonumber
\end{eqnarray} 
where  the values extrapolated from experimental data are given inside the
parentheses. Note that the masses given above are defined in the 
modified minimal subtraction ($\overline{\mbox{MS}}$) scheme and are  
evaluated at $M_{Z}$.
These values correspond to the following set of input parameters
at the GUT scale,  $M_{GUT} = 1.03 \times 10^{16} \; GeV$: 
\begin{eqnarray} 
& a = 0.00250, \quad b =  3.26 \times 10^{-3}\nonumber\\
& c = 0.0346, \quad  d =  0.650\nonumber\\
& \theta  = 0.74 \nonumber\\
& e  = 4.036 \times 10^{-3}, \quad f  =  0.0195 \nonumber\\
& h =  0.06878, \quad \xi  =  -1.52 \nonumber\\
& g_{1} =  g_{2} = g_{3}  =  0.746
\end{eqnarray}
the one-loop renormalization group equations for the MSSM spectrum with three
right-handed neutrinos 
are solved numerically down to the effective
right-handed neutrino mass scale, $M_{R}$. At $M_{R}$, the seesaw mechanism  
is implemented. With the constraints 
$|m_{\nu_{3}}| \gg |m_{\nu_{2}}|, \; |m_{\nu_{1}}|$ and 
maximal mixing in the atmospheric sector, the up-type mass texture leads us 
to choose the following effective neutrino mass matrix
\begin{equation}\label{mll}
M_{\nu_{LL}} = \left(
\begin{array}{ccc}
0 & 0 & t\\
0 & 1 & 1+t^{n}\\
t & 1+t^{n} & 1
\end{array}
\right)\frac{d^{2}v_{u}^{2}}{M_{R}}
\end{equation}
with $n=1.15$, and from the seesaw formula we obtain
\begin{eqnarray}
\label{delta}
\delta_{1} & = & \frac{a^{2}}{r}
\\
\delta_{2} & = & \frac{b^{2}t e^{2i\theta}}
{r}
\\
\delta_{3} & = & \frac{-a(be^{i\theta}(1+t^{1.15})-c)+bct e^{i\theta}}
{r} \; ,
\end{eqnarray}
where $r=(c^{2}t+a^{2}t^{0.15}(2+t^{1.15})-2a(-1+c+ct^{1.15}))$.
We then solve the two-loop RGE's for the MSSM spectrum 
down to the SUSY breaking scale, taken to be $m_{t}(m_{t})=176.4 \; GeV$, and
then the SM RGE's from $m_{t}(m_{t})$ to the weak scale, $M_{Z}$. 
We assume that 
$\tan \beta \equiv v_{u}/v_{d} = 10$,  with 
$v_{u}^{2} + v_{d}^{2} = (246/\sqrt{2} \; GeV) ^{2}$. At the weak scale
$M_{Z}$, the predictions for 
$\alpha_{i}
\equiv g_{i}^{2}/4\pi$ are   
\begin{displaymath} 
\alpha_{1}=0.01663,
\quad \alpha_{2}=0.03374, 
\quad \alpha_{3}=0.1242  \; .
\end{displaymath}
These values compare very well with the values extrapolated to $M_{Z}$ from the
experimental data, 
$(\alpha_{1},\alpha_{2},\alpha_{3})=
(0.01696,0.03371,0.1214 \pm 0.0031)$.
The predictions at the weak scale $M_{Z}$ for the
charged fermion masses, CKM matrix elements and strengths of CP violation, 
are summarized in Table.~\ref{table:predict}. 
\begin{table}
\caption{
The predictions for the charged fermion masses, the 
CKM matrix elements and the CP violation measures.
\label{table:predict}}
\begin{ruledtabular}
\begin{tabular}{l c | c c l c c c l}
 & & & & experimental results \qquad \qquad 
 & & & & predictions at $M_{z}$ \\ 
& & & & extrapolated to $M_{Z}$ \qquad \qquad
 & & & & \\ 
\hline
$m_{s}/m_{d}$  
& & & & $17 \sim 25$  
& & & & $25$\\
$m_{s}$ 
& & & & $93.4^{+11.8}_{-13.0}MeV$  
& & & & $86.0 MeV$\\
$m_{b}$ 
& & & & $3.00\pm 0.11GeV$  
& & & & $3.03 GeV$\\
\hline
$\vert V_{ud} \vert$
& & & & $0.9739-0.9751$ 
& & & & $0.974$\\
$\vert V_{cd} \vert$  
& & & & $0.221-0.227$  
& & & & $0.225$\\
$\vert V_{cs} \vert$ 
& & & & $0.9730-0.9744$  
& & & & $0.973$\\
$\vert V_{td} \vert$  
& & & & $0.0048-0.014$  
& & & & $0.00801$\\
$\vert V_{ts} \vert$  
& & & & $0.037-0.043$ 
& & & & $0.0386$\\
$\vert V_{tb} \vert$  
& & & & $0.9990-0.9992$  
& & & & $0.999$ \\
$J_{CP}^{q}$  
& & & & $(2.88 \pm 0.33) \times 10^{-5}$  
& & & & $2.87 \times 10^{-5}$ \\
$\sin 2\alpha$  
& & & & $-0.16 \pm 0.26$
& & & & $-0.048$ \\
$\sin 2\beta$ 
& & & & $0.736 \pm 0.049$ 
& & & & $0.740$ \\
$\gamma$  
& & & & $60^{0} \pm 14^{0}$
& & & & $64^{0}$\\
$\overline{\rho}$  
& & & & $0.20 \pm 0.09$
& & & & $0.173$\\
$\overline{\eta}$  
& & & & $0.33 \pm 0.05$
& & & & $0.366$
\end{tabular}
\end{ruledtabular}
\end{table}
The predictions of our model in this {\it updated} fit are in 
good agreement with all  
experimental data within $1 \sigma$, including much improved measurements in 
B Physics that give rise to precise values for the CKM matrix elements and for 
the unitarity triangle~\cite{Charles:2004jd}. Note that we have 
taken the SUSY threshold correction to $m_{b}$ to be 
$-18 \%$~\cite{Hall:1993gn}.

The allowed region for the neutrino oscillation parameters 
has been reduced significantly after Neutrino 2004. 
In the atmospheric sector, the global 
analysis including the most recent K2K result yields, 
at $90\%$ CL~\cite{Maltoni:2004ei},
\begin{eqnarray}
 \Delta m_{atm}^{2} =  2.3^{+0.7}_{-0.4} \times 10^{-3} eV^{2}
 \\
 \sin^{2} 2\theta_{atm} >  0.9 \\
 (\mbox{best fit value:}  \sin^{2} 2\theta_{atm} =1.0) \quad .
 \end{eqnarray}
In the solar sector, the global analysis with SNO and most recent KamLAND data 
yields, at $1 \sigma \; (3 \sigma)$~\cite{Bahcall:2004ut},
 \begin{eqnarray}
 \Delta m_{\odot}^{2} & = & 8.2^{+0.3}_{-0.3}(^{+1.0}_{-0.8}) \times 10^{-5} eV^{2}
 \\
 \tan^{2} \theta_{\odot}& = & 0.39^{+0.05}_{-0.04}(^{+0.19}_{-0.11}) \quad .
 \end{eqnarray}
Combining with the CHOOZ result, a global analysis shows that the  
angle $\theta_{13}$ is constrained to be~\cite{Bahcall:2004ut}
\begin{equation}
\sin^{2}\theta_{13} < 0.015 (0.048)
\end{equation}
at $1 \sigma \; (3 \sigma)$.
Using the mass square difference in the atmospheric sector 
$\Delta m_{atm}^{2}=2.33 \times 10^{-3} \; eV^{2}$ and 
the mass square difference for the LMA solution 
$\Delta m_{\odot}^{2}=8.14 \times 10^{-5} \; eV^{2}$ as input 
parameters, we determine 
$t = 0.344$ and $M_{R} = 6.97 \times 10^{12} GeV$, which yield   
$(\delta_{1},\delta_{2},\delta_{3}) 
= (0.00120,0.000703 e^{i \; (1.47)},0.0210 e^{i \;(0.175)})$. We obtain 
the following predictions in the neutrino sector: 
The three mass eigenvalues are give by  
\begin{equation}
(m_{\nu_{1}},m_{\nu_{2}},m_{\nu_{3}}) = (0.00262,0.00939,0.0492) \; eV \; .
\end{equation}
The prediction for the MNS matrix is
\begin{equation}
\vert U_{MNS} \vert = 
\left(
\begin{array}{ccc}
0.852 & 0.511 & 0.116\\
0.427 & 0.560 & 0.710\\
0.304 & 0.652 & 0.695
\end{array}
\right)
\end{equation}
which translates into the mixing angles in the atmospheric, 
solar and reactor sectors,
\begin{eqnarray}
\sin^{2} 2 \theta_{atm} & \equiv & \frac{
4 \vert U_{\mu \nu_{3}} \vert^{2} |U_{\tau \nu_{3}}|^{2}}
{(1-|U_{e\nu_{3}}|^2)^{2}}
= 1.00
\\
\tan^{2} \theta_{\odot} & \equiv & \frac{\vert U_{e \nu_{2}}\vert^{2}}
{|U_{e \nu_{1}}|^{2}} = 0.36
\\
\sin^{2}\theta_{13} & = & |U_{e\nu_{3}}|^{2} = 0.0134 \; .
\end{eqnarray}
The prediction of our model for the strengths of CP violation in 
the lepton sector are
\begin{eqnarray}
J_{CP}^{l} \equiv Im\{ U_{11} U_{12}^{\ast} U_{21}^{\ast} U_{22} \}
= -0.00941
\\
(\alpha_{31},\alpha_{21}) = (0.934,-1.49) \; .
\end{eqnarray}
Using the predictions for the neutrino masses, mixing angles and the 
two Majorana phases, 
$\alpha_{31}$ and $\alpha_{21}$, the matrix element for the neutrinoless double 
$\beta$ decay can be calculated and is given by   
$\vert < m > \vert = 3.1 \times 10^{-3} \; eV$, 
with the present experimental upper bound being $0.35 \; eV$~\cite{Eidelman:2004wy}.
Masses of the heavy right-handed neutrinos are
\begin{eqnarray}
M_{1} & = & 1.09 \times 10^{7} \; GeV
\label{mr1}\\
M_{2} & = & 4.53 \times 10^{9} \; GeV
\label{mr2}\\
M_{3} & = & 6.97 \times 10^{12} \; GeV \; .
\label{mr3}
\end{eqnarray} 
The prediction for the $\sin^{2}\theta_{13}$ value is $0.0134$, 
in agreement with the current bound $0.015$ at $1 \sigma$. 
Because our prediction for 
$\sin^{2}\theta_{13}$ is very close to the present sensitivity 
of the experiment, the validity of our model can be tested in 
the foreseeable future~\cite{Diwan:2003bp}.

\section{Lepton Flavor Violating Decays}\label{lfv}

In light of the neutrino oscillation, extensive searches for lepton flavor 
violation processes, such as $\ell_{i}\rightarrow \ell_{j} \gamma$, 
$\ell_{i}^{-} \rightarrow \ell_{j}^{-} \ell_{j}^{+} \ell_{j}^{-}$, 
muon-electron conversion, are underway. In the SM, as the lepton number is 
conserved, there is no lepton flavor violation. 
Non-zero neutrino masses imply lepton number violation. If neutrino masses are 
induced by the seesaw mechanism, new Yukawa coupling involving the RH neutrinos 
can induce flavor violation~\cite{Masiero:2004js}, 
similar to its quark counter part.
In the non-supersymmetric case, the decay amplitudes for these processes 
are inversely proportional to the RH neutrino mass, $M_{R}^{2}$, 
which is typically much higher than the electroweak scale. 
As a consequence, in non-supersymmetric models, these 
processes are highly suppressed to the level that are unobservable.
 
\begin{figure}[b!]
 \includegraphics[scale=0.55]{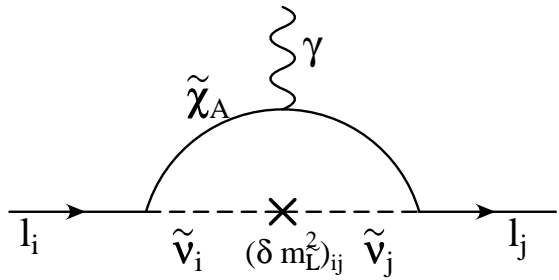}%
 \caption{The dominant diagram that contribute to the decay $\ell_{i} \rightarrow 
\ell_{j} \gamma$ at one loop, mediated by the neutralino $\widetilde{\chi}_{A}$ 
and the sneutrinos $\widetilde{\nu}$. The inserted mass term 
$(\delta m_{\widetilde{L}}^{2})_{ij}$ 
is induced by the renormalization group evolution from the GUT scale to the RH 
neutrino mass scales.
\label{fig.1loop}}
\end{figure}

Significant enhancement in the decay rate can be obtained in supersymmetric models, 
as the characteristic scale in this case is the SUSY scale, which is 
expected to be not too far from the electroweak scale. 
Thus the amplitudes for these decay processes scale as inverse 
square of the SUSY breaking scale, rather than $1/M_{R}^{2}$.  
The relevant interactions that 
give rise to lepton flavor violating decays 
come from the soft-SUSY breaking Lagrangian,
\begin{eqnarray}
-\mathcal{L}_{soft} & = &
(m_{\widetilde{L}}^{2})_{ij} 
\widetilde{\ell}_{L_{i}}^{\dagger} \widetilde{\ell}_{L_{j}} 
+ (m_{\widetilde{e}}^{2})_{ij} 
\widetilde{e}_{R_{i}}^{\dagger} \widetilde{e}_{R_{j}} 
+(m_{\widetilde{\nu}}^{2})_{ij} 
\widetilde{\ell}_{R_{i}}^{\dagger} \widetilde{\ell}_{R_{j}} 
\nonumber\\
& & 
+(\widetilde{m}_{h_{d}}^{2})\widetilde{H}_{d}^{\dagger}\widetilde{H}_{d} 
+ (\widetilde{m}_{h_{2}}^{2}) \widetilde{H}_{u}^{\dagger} \widetilde{H}_{u}
+ \biggl[ A_{\nu}^{ij} \widetilde{H}_{u} 
\widetilde{\nu}_{R_{i}}^{\ast} \widetilde{\nu}_{L_{j}}
\nonumber\\
& &
+  A_{e}^{ij} H_{d} \widetilde{e}_{R_{i}}^{\ast} \widetilde{e}_{L_{j}}
+\frac{1}{2} B_{\nu}^{ij} \widetilde{\nu}_{R_{i}} \widetilde{\nu}_{R_{j}}
+ B_{h}  H_{d} H_{u} 
\nonumber\\
&&+ h.c. \biggr] \quad ,
\end{eqnarray}
where $\widetilde{\ell}_{L}$, $\widetilde{e}_{R}$ 
and $\widetilde{\nu}_{R}$ are the LH 
slepton doublets, RH charged sleptons, and RH sneutrinos, respectively; 
$H_{u}$ ($\widetilde{H}_{u}$) and $H_{d}$ ($\widetilde{H}_{d}$) are 
the two Higgs (higgsino) doublets in MSSM.
Assuming mSUGRA boundary conditions at the GUT scale,
\begin{eqnarray}
(m_{\widetilde{L}}^{2})_{ij} 
= (m_{\widetilde{e}}^{2})_{ij}=(m_{\widetilde{\nu}}^{2})_{ij} 
= m_{0} \delta_{ij}
\\
\widetilde{m}_{H_{d}}^{2} = \widetilde{m}_{H_{u}}^{2} = m_{0}^{2}
\\
A_{\nu}^{ij} = (Y_{\nu})_{ij} A_{0}, \quad A_{e}^{ij} = (Y_{e})_{ij} A_{0}
\\
B_{\nu}^{ij} = M_{\nu_{RR}} B_{0}, \quad B_{h} = \mu B_{0} 
\end{eqnarray}
where $Y_{\nu}$ and $Y_{e}$ are the Yukawa couplings of the neutrinos 
and charged leptons, and $M_{\nu_{RR}}$ is the Majorana mass 
matrix of the RH neutrinos.
As the slepton mass matrix $(m_{\widetilde{L}}^{2})_{ij}$ is 
flavor-blind at the GUT scale, there is no flavor violation 
at $M_{GUT}$. However, as $(m_{\widetilde{L}}^{2})_{ij}$ evolves 
from $M_{GUT}$ to the RH neutrino mass scale, $M_{R}$, 
according to the renormalization group equation,
\begin{eqnarray}
\frac{d}{d\ln \mu} (m_{\widetilde{L}}^{2})_{ij} & = &
\frac{1}{16\pi^{2}} 
\biggl[ 
m_{\widetilde{L}}^{2} (Y_{\nu}^{\dagger}Y_{\nu})_{ij} 
\nonumber\\
& &
+ 2 \biggl( (Y_{\nu}^{\dagger} m_{\widetilde{\nu}}^{2} Y_{\nu})_{ij} 
+ m_{\widetilde{h}}^{2} (Y_{\nu}^{\dagger}Y_{\nu})_{ij} 
\nonumber\\
& & 
+ (A_{\nu}^{\dagger}A_{\nu})_{ij} \biggr)
\biggr] \; , \quad \mbox{for} \; i \ne j \; ,
\end{eqnarray}
the off diagonal elements in the slepton mass matrix $m_{\widetilde{L}}^{2}$ 
can be generated at low energies due to the RG 
corrections~\cite{Hisano:1995cp}, %Hisano:1998fj}, 
\begin{eqnarray}
\delta (m_{\widetilde{L}}^{2})_{ij} & = & -\frac{1}{8\pi} (3m_{0}^{2} + A_{0}^{2})
\nonumber\\
& & \times 
\sum_{k=1,2,3} (\mathcal{Y}_{\nu}^{\dagger})_{ik} (\mathcal{Y}_{\nu})_{kj} 
\ln (\frac{M_{GUT}}{M_{R_{k}}}) \; ,
\end{eqnarray}
for $i \ne j$. 
Here $\mathcal{Y}_{\nu}$ is the Yukawa couplings for the neutrinos in 
the basis where both charged lepton Yukawa matrix and the Majorana 
mass matrix for the RH neutrinos are diagonal; $M_{R_{k}}$ are the 
masses of the heavy neutrinos.
The Yukawa coupling $\mathcal{Y}_{\nu}$ in the new basis is related 
to $Y_{\nu}$ in the original basis by
\begin{equation}\label{ybasis}
\mathcal{Y}_{\nu} = P_{R} O_{R} Y_{\nu} O_{e_{L}}^{\dagger}.
\end{equation} 
Here $O_{L_{e}}$ is the diagonalization matrix for 
\begin{equation}
\mathcal{M}_{e}^{\mbox{\tiny diag}} = O_{e_{R}} M_{e} O_{e_{L}}^{\dagger} \; ,
\end{equation}
and the diagonal phase matrix $P_{R}$ and the orthogonal matrix $O_{R}$ 
are defined by,
\begin{eqnarray}
\mathcal{M}_{\nu_{RR}}^{\mbox{\tiny diag}} & = & 
\mbox{diag}(M_{1}, M_{2},M_{3})
\nonumber\\
& = & P_{R} O_{R} M_{\nu_{RR}} O_{R}^{T} P_{R} \; ,
\end{eqnarray}
where $M_{1,2,3}$ are real and positive, and their numerical values 
are given in Eq.~(\ref{mr1})-(\ref{mr3}).
In our model, the Yukawa matrix $\mathcal{Y}_{\nu}$ is,
\begin{widetext}
\begin{equation}
\mathcal{Y}_{\nu} = 
\left( \begin{array}{ccc}
2.69 \times 10^{-6} e^{-(0.695)i} &
5.92 \times 10^{-5} e^{-(2.75)i} &
6.54 \times 10^{-4} e^{-(1.68)i} \\
1.44 \times 10^{-4} e^{(1.54)i} &
1.73 \times 10^{-3} e^{-(0.176)i} &
8.91 \times 10^{-3} e^{-(1.32)i}\\
2.18 \times 10^{-3} e^{(0.737)i} &
0.0213 e^{(0.0064)i} &
0.618 %e^{5.14\times 10^{-5}}
\end{array}
\right) \;.
\end{equation}
\end{widetext}
The non-vanishing off-diagonal matrix elements in $(\delta m_{\widetilde{L}}^{2})_{ij}$ 
induces lepton flavor violating processes mediated by the superpartners 
of the neutrinos through the one-loop diagram shown in 
Fig.~\ref{fig.1loop}. 
%These processes have been studied before in various 
%models~\cite{Casas:2001sr}. %,Lavignac:2001vp,Ellis:2001xt,Lavignac:2002gf,Ellis:2002xg,Masiero:2002jn,Masiero:2004js}.

In Table \ref{table.lfv} we summarize current status and future 
proposals of the experimental searches for lepton flavor violating decays.
\begin{table*}
 \caption{Summary of current status and future proposals of the 
  experimental searches for lepton flavor violating decays.
 \label{table.lfv}}
 \begin{ruledtabular}
 \begin{tabular}{lll}
 Decay & current bound on the branching ratio & reach of future experiment\\
 \hline
 $\mu \rightarrow e\gamma$ &
 $< 1.2 \times 10^{-11}$ (MEGA, 1999)\cite{Brooks:1999pu} &
 $10^{-14}$ (PSI)\cite{Barkov:1999}\\
 & & $10^{-15}$ (J-PARC)\\
 $\mu \rightarrow 3e$ &
 $< 1.0 \times 10^{-12}$ (SINDRUM, 1988)\cite{Bellgardt:1987du} &
 \\
 $\mu \rightarrow e$ in ${ }^{48}_{22}Ti$&
 $< 6.1\times 10^{-13}$ (SINDRUM II, 1998)\cite{Wints:1998} &
 $2.0 \times 10^{-17}$ (MECO)\cite{Bachman:1997}\\
 & & $10^{-18}$ (J-PARC) \\
 $\tau \rightarrow \mu \gamma$ &
 $< 3.1\times 10^{-7}$ (BELLE, 2003) \cite{Inami:2003}&
 $10^{-9}$ (BELLE)\cite{Inami:2003}\\
 $\tau \rightarrow e\gamma$ &
 $< 3.6 \times 10^{-7}$ (BELLE, 2003) \cite{Abe:2003sx} &
 \end{tabular}
 \end{ruledtabular}
 \end{table*}
In the following subsections, we discuss each LFV process individually. 
%in the context of our model.

\subsection{$\mu \rightarrow e \gamma$, $\tau \rightarrow 
\mu \gamma$ and $\tau \rightarrow e\gamma$}

The branching ratios for the decay of $\ell_{i} \rightarrow \ell_{j} + \gamma$ 
induced by the renormalization group effects described above 
is given by~\cite{Hisano:1995cp}
\begin{eqnarray}
Br(\ell_{i} \rightarrow \ell_{j} \gamma) 
& = & \frac{\alpha^{3}}{G_{F}^{2}m_{S}^{8}} 
|\frac{-1}{8\pi} (3m_{0}^{2} + A_{0}^{2}) |^{2} \tan^{2} \beta
\nonumber\\
& &  \times
|\sum_{k=1,2,3} (\mathcal{Y}_{\nu}^{\dagger})_{ik} (\mathcal{Y}_{\nu})_{kj} 
\ln (\frac{M_{GUT}}{M_{R_{k}}}) |^{2}
\quad .\nonumber\\
\end{eqnarray}
Here $\alpha$ is the fine structure constant, $G_{F}$ is the 
Fermi constant, 
and $m_{S}$ is the typical SUSY scalar mass which is given by, 
to a very good approximation~\cite{Petcov:2003zb}, 
\begin{equation}
m_{S}^{8} = \frac{1}{2} m_{0}^{2} M_{1/2}^{2} (m_{0}^{2} + 0.6 M_{1/2}^{2})^{2},
\end{equation}
where $M_{1/2}$ is the universal gaugino mass. 
In our model, $|\delta (m_{\widetilde{L}}^{2})_{ij}|$ is given by,
\begin{eqnarray}
|\delta (m_{\widetilde{L}}^{2})_{ij} |
& = & |\frac{1}{8\pi} (3m_{0}^{2} + A_{0}^{2})| 
\nonumber\\
& & \times
\left(\begin{array}{ccc}
\ast & 3.41 \times 10^{-4} & 0.0098\\
3.41 \times 10^{-4} & \ast & 0.0962\\
0.0098 & 0.0962 & \ast
\end{array}\right) \; , \nonumber\\
\end{eqnarray}
for $i \ne j$. 
Thus the following relation is predicted,
\begin{equation}
Br(\mu \rightarrow e \gamma) < Br(\tau \rightarrow e \gamma) 
< Br(\tau \rightarrow \mu \gamma) \; .
\end{equation}
Similar relation was observed in Ref.~\cite{Bando:2004hi} in which symmetric 
mass matrices with four texture zeros are utilized. We also note that 
the value for $\tan\beta$ is $10$, thus there is no $\tan\beta$ enhancement 
in our predictions.

\begin{figure}[t!]
 \includegraphics[scale=0.55,angle=270]{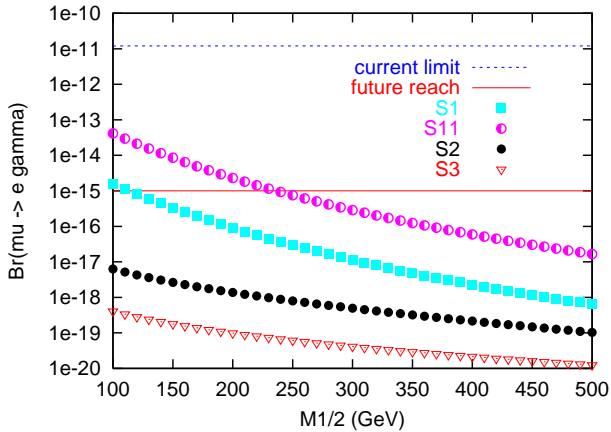}%
 \caption{The branching ratio of the decay $\mu \rightarrow e \gamma$ 
as a function of the universal gaugino mass $M_{1/2}$ for various  
scalar masses $A_{0}$ and $m_{0}$. (S1): $m_{0} = A_{0} = 100 \; GeV$; 
(S11): $m_{0} = 100 \; GeV, A_{0} = 1 \; TeV$; (S2): $m_{0} = A_{0} = 500 \; GeV$;
(S3): $m_{0} = A_{0} = 1 \; TeV$. The dash line corresponds to the current 
experimental limit $1.2 \times 10^{-11}$ from MEGA, while the solid line indicates 
the reach of a future experiment at J-PARC, $10^{-15}$. The value of $\tan\beta$ 
in our model is $\tan\beta = 10$. 
  \label{fig.muegamma}}
\end{figure}

\begin{figure}[t!]
 \includegraphics[scale=0.55,angle=270]{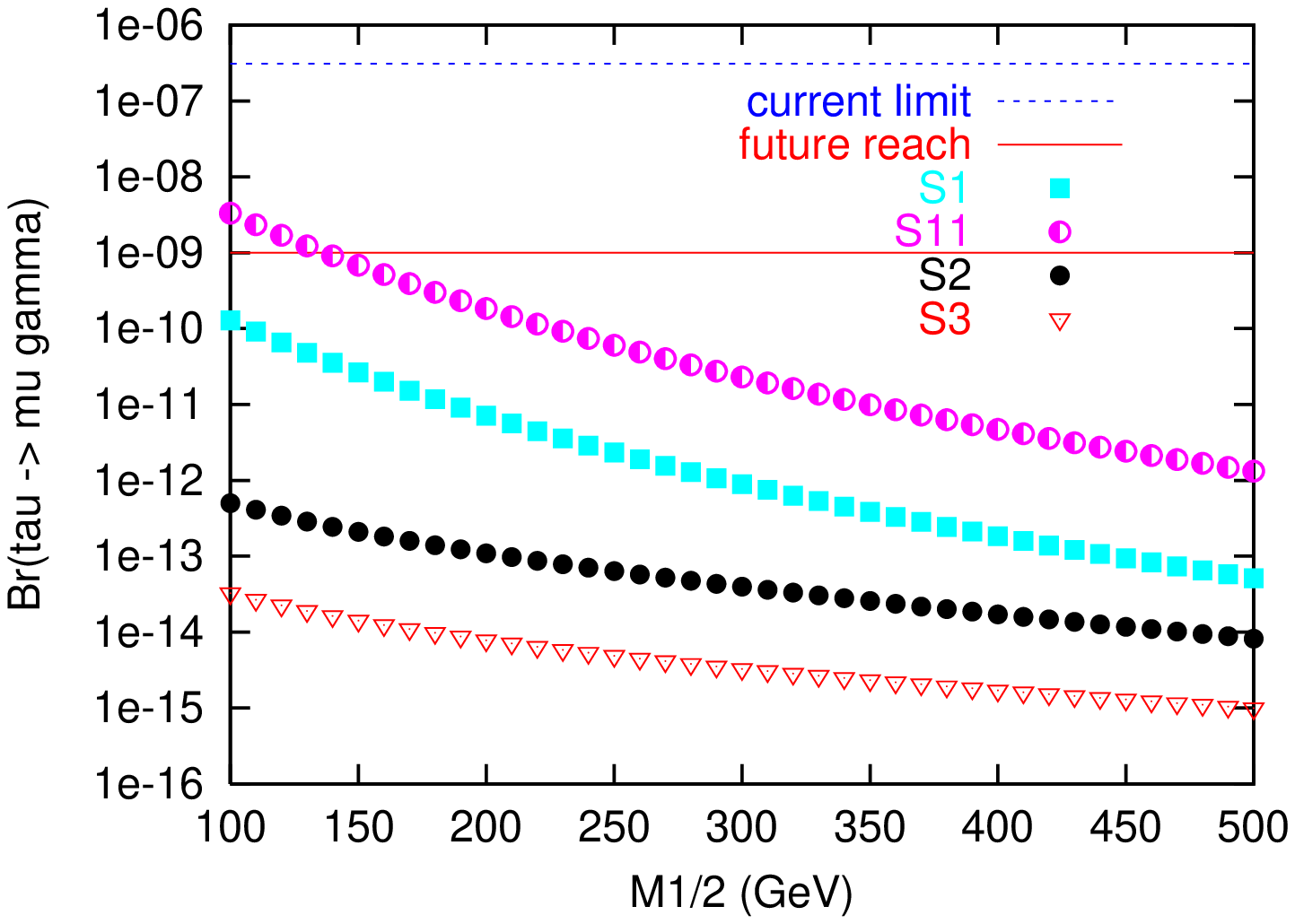}%
 \caption{The branching ratio of the decay $\tau \rightarrow \mu \gamma$ 
as a function of the universal gaugino mass $M_{1/2}$ for various  
scalar masses $A_{0}$ and $m_{0}$. (S1): $m_{0} = A_{0} = 100 \; GeV$; 
(S11): $m_{0} = 100 \; GeV, A_{0} = 1 \; TeV$; (S2): $m_{0} = A_{0} = 500 \; GeV$;
(S3): $m_{0} = A_{0} = 1 \; TeV$. The dash line corresponds to the current 
experimental limit $3.1 \times 10^{-7}$ from BELLE, while the solid line indicates 
the reach of a future experiment at BELLE, $10^{-9}$. The value of $\tan\beta$ 
in our model is $\tan\beta = 10$.   
\label{fig.taumugamma}}
\end{figure}

\begin{figure}[t!]
 \includegraphics[scale=0.55,angle=270]{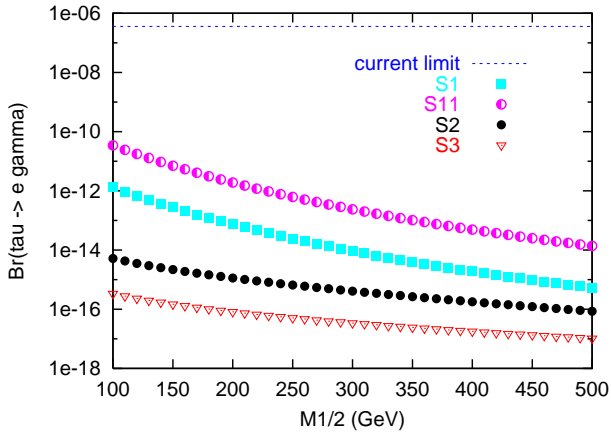}%
 \caption{The branching ratio of the decay $\tau \rightarrow e \gamma$ 
as a function of the universal gaugino mass $M_{1/2}$ for various  
scalar masses $A_{0}$ and $m_{0}$. (S1): $m_{0} = A_{0} = 100 \; GeV$; 
(S11): $m_{0} = 100 \; GeV, A_{0} = 1 \; TeV$; (S2): $m_{0} = A_{0} = 500 \; GeV$;
(S3): $m_{0} = A_{0} = 1 \; TeV$. The dash line corresponds to 
the current upper bound, $3.6 \times 10^{-7}$, from BELLE. The value of $\tan\beta$ 
in our model is $\tan\beta = 10$. \label{fig.tauegamma}}
\end{figure}

Currently the most stringent experimental  bound on the lepton 
flavor violating processes is on the decay $\mu \rightarrow e \gamma$. 
The prediction of our model for  $Br(\mu \rightarrow e \gamma)$ is well 
below the most stringent bound up-to-date from MEGA at LANL~\cite{Brooks:1999pu}. 
In Fig. \ref{fig.muegamma}, the branching ratio of the decay 
$\mu \rightarrow e \gamma$ as a function of the universal gaugino mass 
$M_{1/2}$ is shown for various scalar masses $A_{0}$ and $m_{0}$. 
For large $A_{0}$ and low $m_{0}$ and $M_{1/2}$, 
there is a large soft SUSY parameter space that give rise 
to predictions which can be probed by MEG at PSI and/or at J-PARC. 
In Fig. \ref{fig.taumugamma}, the branching ratio of the decay 
$\tau \rightarrow \mu \gamma$ as a function of the universal gaugino mass 
$M_{1/2}$ is shown for various scalar masses $A_{0}$ and $m_{0}$. 
For $A_{0} \sim \mathcal{O}(1 \; TeV)$ and $m_{0}$ and $M_{1/2}$ 
both of order $\mathcal{O}(100 \; GeV)$, the prediction of 
our model on $\tau \rightarrow \mu \gamma$ may be tested 
at BELLE in the future.
In Fig. \ref{fig.tauegamma}, the branching ratio of the decay 
$\tau \rightarrow e \gamma$ as a function of the universal  gaugino mass 
$M_{1/2}$ is shown for various scalar mass $A_{0}$ and $m_{0}$. 
For the SUSY parameter space we consider, the prediction
for $Br(\tau \rightarrow e \gamma)$ is at least four orders of 
magnitudes below the current experimental upper bound.

We comment that, in models with lop-sided textures~\cite{Albright:1998vf}, 
the maximal mixing angle observed in the atmospheric neutrino sector is due to 
a large $(23)$ mixing in the charged lepton sector. 
As a result, the off-diagonal elements in $(23)$ sector of $O_{e_{L}}$ are of order 
$\mathcal{O}(1)$, which in turn gives rise to an enhancement in the decay 
branching ratios. In order to satisfy the current experimental 
upper bound, some new mechanism must be in place to suppress the decay rate of 
$\mu \rightarrow e\gamma$ in models with lop-sided textures~\cite{Barr:2003fn}. 
In our model which utilizes symmetric textures, as
large leptonic mixing in our model is a result of the seesaw mechanism, 
all off-diagonal matrix elements in $Y_{\nu}$, $O_{e_{L}}$ and $O_{R}$ are 
much smaller than unity, leading to a much smaller branching ratio for 
$\mu \rightarrow e\gamma$ than that predicted in models with lop-sided textures.
Yet our prediction is large enough to be probed by the next generation of experiments 
within a few years. 

\subsection{$\mu \rightarrow 3e$}

\begin{figure}[t!]
 \includegraphics[scale=0.55,angle=270]{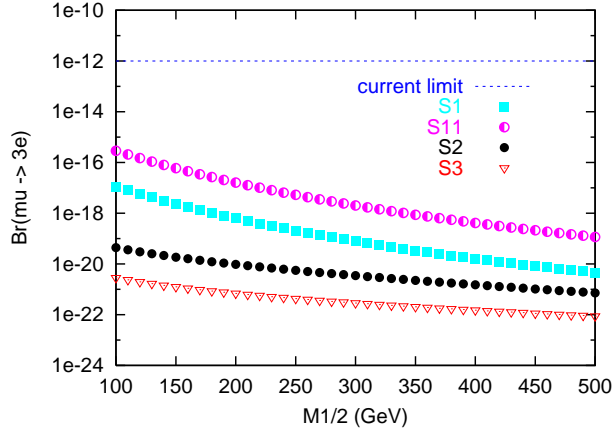}%
 \caption{The branching ratio of the decay $\mu^{-} \rightarrow e^{-} e^{+} e^{-}$ 
as a function of the universal gaugino mass $M_{1/2}$ for various  
scalar masses $A_{0}$ and $m_{0}$. (S1): $m_{0} = A_{0} = 100 \; GeV$; 
(S11): $m_{0} = 100 \; GeV, A_{0} = 1 \; TeV$; (S2): $m_{0} = A_{0} = 500 \; GeV$;
(S3): $m_{0} = A_{0} = 1 \; TeV$. The dash line corresponds to the current 
experimental limit $1.0 \times 10^{-12}$ from SINDRUM. The value of $\tan\beta$ 
in our model is $\tan\beta = 10$.  \label{fig.mu3e}}
\end{figure}

For the process $\mu \rightarrow 3e$, as penguium diagrams are the dominant contributions, 
the branching ratio of the decay $\ell_{i}^{-} 
\rightarrow \ell_{j}^{-} \ell_{j}^{+} 
\ell_{j}^{-}$ has similar structure as that of  
the decay $\ell_{i}^{-} \rightarrow \ell_{j}^{-}  \gamma$. 
To a very good approximation, the relation between these two processes 
reads~\cite{Hisano:1995cp}, 
\begin{equation}
\frac{Br(\ell_{i}^{-} \rightarrow 
\ell_{j}^{-} \ell_{j}^{+} \ell_{j}^{-})}
{Br(\ell_{i}^{-} \rightarrow \ell_{j}^{-}  \gamma)}
\simeq \frac{\alpha}{8\pi} \biggl[ \frac{16}{3} 
\ln(\frac{m_{\ell_{i}}}{2m_{\ell_{j}}}) 
- \frac{14}{9} \biggr] 
\quad ,
\end{equation}
where $m_{\ell_{i}}$ is the $i-$th generation lepton mass. 
For the decay $\mu \rightarrow 3e$, we thus have 
\begin{equation}
Br(\mu \rightarrow 3e) \simeq 7 \times 10^{-3} Br(\mu \rightarrow e \gamma) \; .
\end{equation}
In Fig.~\ref{fig.mu3e}, the branching ratio of the decay 
$\mu \rightarrow 3e$ as a function of the universal gaugino mass 
$M_{1/2}$ is shown for various scalar mass $A_{0}$ and $m_{0}$. 
As the current experimental upper bound and the reach of the next phase of 
experiment at BELLE are still quite high, the prediction for $\mu \rightarrow 3e$ 
in our model can not be tested, even with a high value of the scalar mass, 
$A_{0} = 1 \; TeV$. 

\subsection{$\mu$-$e$ Conversion}

\begin{figure}[t!]
 \includegraphics[scale=0.55,angle=270]{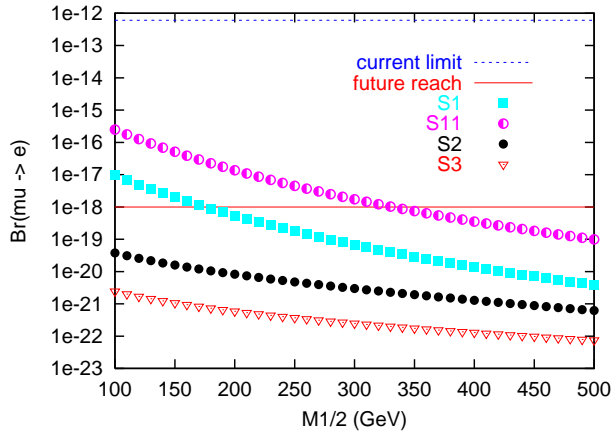}%
 \caption{The branching ratio of the decay 
$\mu^{-} \rightarrow e^{-}$ in ${ }^{48}_{22}$Ti. 
as a function of the universal gaugino mass $M_{1/2}$ for various  
scalar masses $A_{0}$ and $m_{0}$. (S1): $m_{0} = A_{0} = 100 \; GeV$; 
(S11): $m_{0} = 100 \; GeV, A_{0} = 1 \; TeV$; (S2): $m_{0} = A_{0} = 500 \; GeV$;
(S3): $m_{0} = A_{0} = 1 \; TeV$. The dash line corresponds to the current 
experimental limit $6.1 \times 10^{-13}$ from SINDRUM II, 
while the solid line indicates 
the reach of a future experiment at J-PARC, $10^{-18}$.  The value of $\tan\beta$ 
in our model is $\tan\beta = 10$. \label{fig.mue}}
\end{figure}

Similar to the case of $\mu \rightarrow 3e$, 
the branching ratio for muon-electron conversion is also related to 
the branching ratio of the decay $\mu \rightarrow e\gamma$ as long as $\tan\beta$ 
is not too small. In the region $\tan\beta > 1$, the relation between these 
two processes is given by, 
to a very good approximation~\cite{Hisano:1995cp},
\begin{equation}
\frac{Br(\mu \rightarrow e)}{Br(\mu \rightarrow e\gamma)} 
\simeq 16\alpha^{4} Z_{eff}^{4} Z |F(q^{2})|^{2} \quad ,
\end{equation}
where $Z_{eff}$ is the effective charge of the nucleon, $Z$ is the 
proton number and $F(q^{2})$ is the nuclear 
form factor at momentum transfer $q$. For ${ }^{48}_{22}Ti$, the conversion rate is  
\begin{equation}
Br(\mu \rightarrow e; \; { }^{48}_{22}Ti) \simeq 6 \times 10^{-3} 
Br(\mu \rightarrow e\gamma) \quad,
\end{equation}
where $Z_{eff} = 17.6$ and $F(q^{2} = -m_{\mu}^{2}) = 0.54$ have been used.
In Fig. \ref{fig.mue}, the branching ratio of the decay 
$\mu \rightarrow e$ in ${ }^{48}_{22}Ti$ as a function of the universal gaugino mass 
$M_{1/2}$ is shown for various scalar mass $A_{0}$ and $m_{0}$. 
For low values of $m_{0}$ and $M_{1/2}$, 
there is a very large soft SUSY parameter space that give 
rise to prediction for $\mu-e$ conversion rate that is sensitive to 
MECO~\cite{Bachman:1997} at BNL and the proposal at J-PARC.

\section{Baryogenesis {\it \`{a} la}  Soft Leptogenesis}\label{leptogenesis}

It is well known that the CP violation in the quark sector is too small 
to explain the observed baryon asymmetry of the Universe (BAU), expressed 
in terms of the ratio of the baryon number to 
entropy~\cite{Bennett:2003bz},
\begin{equation}  
\frac{n_{b}}{ s} = (0.87 \pm 0.04) \times 10^{-10} \; , 
\end{equation}
derived from CMB and nucleosynthesis measurements.
In leptogenesis, leptonic CP violating phases are used to 
produce asymmetry in leptonic number which
then is converted into baryon asymmetry by the electroweak
non-perturbative effects due to sphalerons. 
There are two ways of producing lepton number
asymmetry: (i) Standard leptogenesis 
(STDL)~\cite{Fukugita:1986hr} %,Luty:1992un,Buchmuller:1996pa}
and (ii) Soft leptogenesis 
(SFTL)~\cite{Grossman:2003jv,D'Ambrosio:2003wy,Boubekeur:2002jn}.

In STDL scenario, the primordial leptonic 
asymmetry is generated by the decay of the heavy right-handed Majorana neutrinos and
their scalar partners, mediated by the Yukawa interactions in the superpotential. 
In our model, the large hierarchy among the three heavy neutrinos leads to 
a very small CP asymmetry, which is of the order of $\mathcal{O}(10^{-9})$.
In addition, the low value for the mass of the lightest RH neutrino, 
$M_{1} = 1.09 \times 10^{7} \; GeV$, leads to an extremely large wash-out effect.  
Due to these reasons, the prediction in our model for the baryonic asymmetry 
utilizing the standard leptogenesis is of the order of $\mathcal{O}(10^{-15})$, 
which is four orders of magnitude below the value derived from experimental observations. 

SFTL utilizes the soft SUSY breaking sector,  
and the asymmetry in the lepton number is 
generated in the decay of the superpartner of the RH 
neutrinos~\cite{Grossman:2003jv,D'Ambrosio:2003wy},
 as opposed to the lightest RH neutrino in the case of STDL.
Unlike in STDL where the Yukawa sector is responsible 
for the required CP violation and lepton number violation,  
in the scenario of SFTL, the CP violation 
and lepton number violation 
trace their origins to SUSY breaking.
As a result, it allows a much lower bound on 
the mass of the lightest RH neutrino, $M_{1}$, 
compared to that in STDL. 
In fact, it has been shown 
very recently that in contrast to the STDL scenario in which 
$M_{1} > 10^{9} GeV$ is typically required to have sufficient baryonic 
asymmetry~\cite{Buchmuller:2002jk}, 
SFTL can only work in the region where 
$M_{1} < 10^{9} \; GeV$~\cite{Grossman:2004dz}. 
As a result, the problem of the gravitino 
over-production~\cite{Ross:1995dq} may be avoided. 

For SFTL, the relevant soft SUSY Lagrangian  
that involves lightest RH sneutrinos $\widetilde{\nu}_{R_{1}}$ 
is the following,
\begin{eqnarray}
-\mathcal{L}_{soft} & = & (\frac{1}{2} B M_{1} \widetilde{\nu}_{R_{1}} 
\widetilde{\nu}_{R_{1}} 
+ A \mathcal{Y}_{1i} \widetilde{L}_{i} \widetilde{\nu}_{R_{1}} H_{u} + h.c.) 
\nonumber\\
& & \quad + \widetilde{m}^{2} \widetilde{\nu}_{R_{1}}^{\dagger} 
\widetilde{\nu}_{R_{1}}
\; .
\end{eqnarray}
This soft SUSY Lagrangian and the superpotential that involves 
the lightest RH neutrino, $N_{1}$, 
\begin{equation}
W = M_{1} N_{1} N_{1} + \mathcal{Y}_{1i} L_{i} N_{1} H_{u}
\end{equation}
give rise to the following interactions 
\begin{eqnarray}
-\mathcal{L}_{\mathscr{A}} & = & 
\widetilde{\nu}_{R_{1}} (
M_{1} Y_{1i}^{\ast} \widetilde{\ell}_{i}^{\ast} H_{u}^{\ast}
+\mathcal{Y}_{1i} \overline{\widetilde{H}}_{u} \ell_{L}^{i} 
+ A \mathcal{Y}_{1i} \widetilde{\ell}_{i} H_{u}
) 
\nonumber\\
& & \qquad + h. c. \quad ,
\end{eqnarray}
and mass terms (to leading order in soft SUSY breaking terms),
\begin{equation}
-\mathcal{L}_{\mathscr{M}}  =  
(M_{1}^{2} \widetilde{\nu}_{R_{1}}^{\dagger} \widetilde{\nu}_{R_{1}}  +  
\frac{1}{2} B M_{1} 
\widetilde{\nu}_{R_{1}} \widetilde{\nu}_{R_{1}} ) + h.c. \; .
\end{equation}
Diagonalization of the mass matrix $\mathcal{M}$ with 
the two states $\widetilde{\nu}_{R_{1}}$ and 
$\widetilde{\nu}_{R_{1}}^{\dagger}$ 
leads to eigenstates  
$\widetilde{N}_{+}$ and $\widetilde{N}_{-}$ 
with masses,  
\begin{equation}
M_{\pm} \simeq M_{1} (1 \pm \frac{|B|}{2M_{1}}) \; ,
\end{equation}
where the leading order term $M_{1}$ is the F-term contribution 
from the superpotential (RH neutrino mass term) and    
the mass difference between the two mass eigenstates $\widetilde{N}_{+}$ 
and $\widetilde{N}_{-}$ is induced by the SUSY breaking $B$ term.
The time evolution of the 
$\widetilde{\nu}_{R_{1}}$-$\widetilde{\nu}^{\dagger}_{R_{1}}$ system  
is governed by the Schr\"{o}dinger equation, 
\begin{equation}
\frac{d}{dt} \left(
\begin{array}{c}
\widetilde{\nu}_{R_{1}}\\
\widetilde{\nu}_{R_{1}}^{\dagger}
\end{array}\right)
= \mathcal{H}
\left(
\begin{array}{c}
\widetilde{\nu}_{R_{1}}\\
\widetilde{\nu}_{R_{1}}^{\dagger}
\end{array}\right) \; ,
\end{equation}
where the Hamiltonian $\mathcal{H}$ is given 
by~\cite{Grossman:2003jv,D'Ambrosio:2003wy}, 
\begin{eqnarray}
\mathcal{H} & = & \mathcal{M} - \frac{i}{2} \mathscr{A}
\\
\mathcal{M} & = & \left(
\begin{array}{cc}
1 & \frac{B^{\ast}}{2M_{1}}\\
\frac{B}{2M_{1}} & 1
\end{array}\right) \; M_{1} \; ,
\\
\mathscr{A} & = & \left(
\begin{array}{cc}
1 & \frac{A^{\ast}}{M_{1}}\\
\frac{A}{M_{1}} & 1
\end{array}\right) \Gamma_{1} \; .
\end{eqnarray}
For the decay of the lightest RH sneutrino, $\widetilde{\nu}_{R_{1}}$, 
the total decay width  $\Gamma_{1}$ is given by, in the basis 
defined in Eq.~(\ref{ybasis}) where both the charged lepton 
mass matrix and the RH neutrino mass matrix are diagonal,
\begin{equation}
\Gamma_{1} =  \frac{1}{4\pi} 
(\mathcal{Y}_{\nu}\mathcal{Y}_{\nu}^{\dagger})_{11} M_{1}
= 0.374 \; GeV \; .
\end{equation}
The eigenstates of the Hamiltonian $\mathcal{H}$ are  
$\widetilde{N}_{\pm}^{\prime} = p \widetilde{N} 
\pm q \widetilde{N}^{\dagger}$, where $|p|^{2} + |q|^{2} = 1$.
The ratio $q/p$ is given in terms of $\mathcal{M}$ and $\Gamma$ as,
\begin{eqnarray}
\biggl( \frac{q}{p} \biggr)^{2} & = & 
\frac{2\mathcal{M}_{12}^{\ast} - i \mathscr{A}_{12}^{\ast}}
{2\mathcal{M}_{12} - i \mathscr{A}_{12}}
\nonumber\\
& \simeq & 
1 + Im \biggl( \frac{2\Gamma_{1} A}{BM_{1}} \biggr) \; ,
\end{eqnarray}
in the limit $\mathscr{A}_{12} \ll \mathcal{M}_{12}$.
Similar to the $K^{0}-\overline{K}^{0}$ system, 
the source of CP violation in the lepton number asymmetry 
considered here is due to 
the CP violation in the mixing which occurs when the two neutral 
mass eigenstates ($\widetilde{N}_{+}$, $\widetilde{N}_{-}$), 
are different from the interaction eigenstates,
($\widetilde{N}^{\prime}_{+}$, $\widetilde{N}^{\prime}_{-}$).
Therefore CP violation in mixing is present
as long as the quantity $|q/p| \ne 1$, which requires 
\begin{equation}
Im \biggl( \frac{A\Gamma_{1}}{M_{1} B} \biggr) \ne 0 \; . 
\end{equation} 
For this to occur, SUSY breaking, {\it i.e.} non-vanishing $A$ 
{\it and} $B$, is required.
As the relative phase between the parameters $A$ and $B$ 
can be rotated away by an $U(1)_{R}$-rotaion, without loss of generality  
we assume from now on that the physical phase that remains 
is solely coming from the tri-linear coupling $A$. 

The total lepton number asymmetry integrated over time, $\epsilon$,  
is defined as the ratio of difference to the sum of the decay widths $\Gamma$ 
for $\widetilde{\nu}_{R_{1}}$ and $\widetilde{\nu}_{R_{1}}^{\dagger}$ 
into final states of the slepton doublet $\widetilde{L}$ and the Higgs doublet $H$, 
or the lepton doublet $L$ and the higgsino $\widetilde{H}$ or their conjugates,
\begin{equation}
\epsilon = \frac{\sum_{f} \int_{0}^{\infty} [
\Gamma(\widetilde{\nu}_{R_{1}}, \widetilde{\nu}_{R_{1}}^{\dagger} \rightarrow
f) - 
\Gamma(\widetilde{\nu}_{R_{1}}, \widetilde{\nu}_{R_{1}}^{\dagger} \rightarrow
\overline{f})]}
{\sum_{f} \int_{0}^{\infty} 
[\Gamma( \widetilde{\nu}_{R_{1}}, \widetilde{\nu}_{R_{1}}^{\dagger} 
\rightarrow f) + 
\Gamma(\widetilde{\nu}_{R_{1}}, \widetilde{\nu}_{R_{1}}^{\dagger}
\rightarrow \overline{f})]}
\end{equation}
where final states $f = (\widetilde{L}\; H), 
\; (L \; \widetilde{H})$ have lepton number
$+1$, and $\overline{f}$ denotes their conjugate, $(\widetilde{L}^{\dagger} 
\; H^{\dagger}), 
\; (\overline{L} \; \overline{\widetilde{H}})$, which have lepton number
$-1$.
After carrying out the time integration, the total CP asymmetry 
is~\cite{Grossman:2003jv,D'Ambrosio:2003wy}, 
\begin{equation}
\epsilon = \biggl(
\frac{4\Gamma_{1} B}{\Gamma_{1}^{2}+4B^{2}} \biggr)
\frac{Im(A)}{M_{1}} \delta_{B-F}
\end{equation}
where the additional factor $\delta_{B-F}$ takes into account the thermal effects 
due to the difference between the occupation numbers of bosons and 
fermions~\cite{Covi:1997dr}. 
The final result for the baryon asymmetry is~\cite{Grossman:2003jv,D'Ambrosio:2003wy},
\begin{eqnarray}
\frac{n_{B}}{s} & \simeq & 
- c \; d_{\widetilde{\nu}_{R}} \; 
\epsilon \; \kappa
\nonumber\\
& \simeq & 
-1.48 \times 10^{-3} \epsilon \; \kappa
\nonumber\\
& \simeq & -(1.48 \times 10^{-3}) 
\biggl( \frac{Im(A)}{M_{1}} \biggl)
\; R \; \delta_{B-F} \; \kappa
\end{eqnarray}
where $d_{\widetilde{N}}$ in the first line is the density of the lightest sneutrino 
in equilibrium in units of entropy density, and is given by, $d_{\widetilde{\nu}_{R}} 
= 45 \zeta(3)/(\pi^{4}g_{\ast})$; the factor $c = (8N_{F}+4N_{H})/(22N_{F}+13N_{H})$ 
characterizes the amount of $B-L$ asymmetry being converted into the baryon asymmetry $Y_{B}$, 
with $N_{F}$ and $N_{H}$ being the number of families and the  
$SU(2)$ Higgs doublets, respectively. 
For the MSSM particle spectrum, $(N_{F},N_{H})=(3,2)$. The parameter 
$\kappa$ is the dilution factor which  
characterizes the wash-out effects due to the inverse decays and lepton number violating
scattering processes together with the time evolution of the system. 
It is obtained by solving the Boltzmann equations for the system. 
An approximation is given by~\cite{Kolb:1990vq}
\begin{eqnarray}
10^{6} \le r: & \quad 
\kappa = (0.1 r)^{1/2} e^{-(\frac{4}{3})(0.1r)^{1/4}} \\
10 \le r \le 10^{6}: & \quad \kappa = 0.3/(r (\ln r)^{0.6}) \\
0 \le r \le 10: & \quad \kappa = 1/(2\sqrt{r^{2}+9}) \; . 
\end{eqnarray}
where $r$ is defined as
\begin{equation}
r \equiv \frac{M_{pl}}{(1.7)(32\pi)\sqrt{g_{\ast}}} 
\frac{( \mathcal{Y}_{\nu} \mathcal{Y}_{\nu}^{\dagger} )_{11}}{M_{1}}
\end{equation}
with $M_{Pl}$ being the Planck scale taken to be $1.2 \times 10^{19} \; GeV$. 
We have $r = 183$ and correspondingly $\kappa = 0.00061$ in our model.  
The parameter $R$ is defined as the ratio, 
\begin{equation}
R \equiv \frac{4 \Gamma_{1} B}{\Gamma_{1}^{2} + 4 B^{2}} \; ,
\end{equation}
which gives a value equal to one when the resonance condition, $\Gamma_{1} = 2|B|$, 
is satisfied,  
leading to maximal CP asymmetry. As $\Gamma_{1}$ is of the order of 
$\mathcal{O}(0.1-1) \; GeV$, to satisfy the resonance condition, 
a small value for $B \ll \widetilde{m}$ is thus needed. 
Such a small value of $B$ can be generated by some dynamical 
relaxation mechanisms~\cite{Yamaguchi:2002zy} 
in which $B$ vanishes in the leading order. A small 
value of $B \sim \widetilde{m}^{2}/M_{1}$ is then generated by an operator 
$\int d^{4} \theta ZZ^{\dagger}N_{1}^{2} / M_{pl}^{2}$ in the K\"{a}hler 
potential, where $Z$ is the SUSY breaking spurion field, 
$Z = \theta^{2}~\widetilde{m} M_{pl}$~\cite{D'Ambrosio:2003wy}.
In our model, with the parameter $B^\prime  \equiv \sqrt{BM_{1}}$ 
having the size of the natural 
SUSY breaking scale $\sqrt{\widetilde{m}^{2}} \sim \mathcal{O}(1 \; TeV)$, a small 
value for $B$ required by the resonance condition 
$B \sim \Gamma_{1} \sim \mathcal{O}(0.1 \; GeV)$ can thus be obtained. 

\begin{figure}[b]
 \includegraphics[scale=0.55,angle=270]{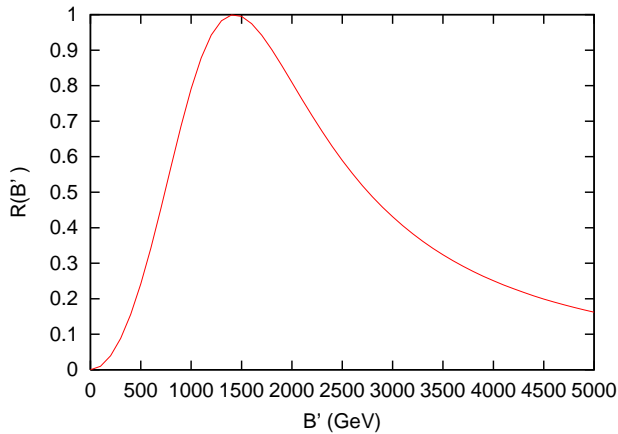}%
 \caption{The ratio $R$ as a function of $B^\prime$. The resonance occurs at around 
$B^\prime \sim 1.4 \; TeV$.
\label{softlpg.res}}
\end{figure}

\begin{figure}[t]
 \includegraphics[scale=0.55,angle=270]{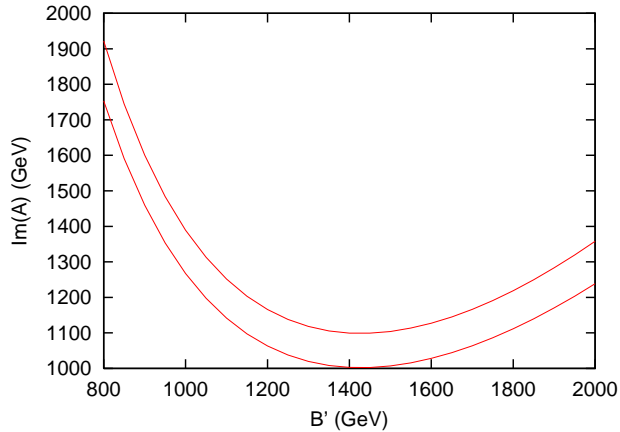}%
 \caption{
The parameter space on the $Im(A)$ versus $B^{\prime}$ plane that gives rise to 
an amount of baryon asymmetry consistent with the value derived from observations, 
$n_{B}/s = (0.87 \pm 0.04) \times 10^{-10}$,  
is the region bounded by these two curves. 
The upper curve corresponds to the upper bound from observation, 
$n_{B}/s = 0.91 \times 10^{-10}$, while the lower curve corresponds to 
the lower bound, $n_{B}/s = 0.83 \times 10^{-10}$.
 \label{softlpg.ab}}
\end{figure}

\begin{figure}[t]
 \includegraphics[scale=0.55,angle=270]{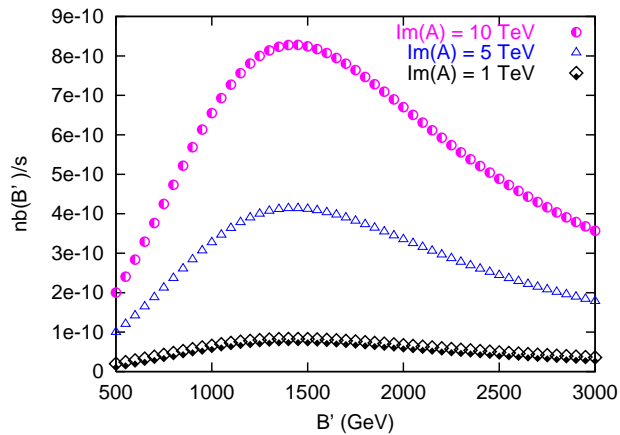}%
 \caption{The prediction for $n_{B}/s$ as a function of $B^\prime$ for 
$|Im(A)| = 10 \; TeV$ (circles), $5 \; TeV$ (triangles) and 
$1 \; TeV$ (squares).  
\label{softlpg.nb}}
\end{figure}

Fig.~\ref{softlpg.res} shows the ratio $R$ 
as a function of $B^\prime$. For the specific value of the decay width 
$\Gamma_{1}$ predicted in our model, the resonance occurs at around 
$B^\prime \sim 1.4 \; TeV$.
In Fig.~\ref{softlpg.ab}, the  
region on the $Im(A)$ versus $B^{\prime}$ plane that gives rise to an  
amount of baryon asymmetry consistent with the value derived from 
observation,  $n_{B}/s = (0.87 \pm 0.04) \times 10^{-10}$, is shown. 
The required value for $B^\prime$ near the resonance is 
around $800 \; GeV - 2 \; TeV$, and the required 
value for $|Im(A)|$ is around $(1-2) \; TeV$. At the resonance $B^\prime$, 
the value for $|Im(A)|$ can be as low as $~ 1 \; TeV$ to generate sufficient 
amount of baryon asymmetry. In Fig.~\ref{softlpg.nb}, we show the predictions 
for the asymmetry, $n_{B}/s$, as a function of $B^\prime$ for different values 
of $Im(A)$. 
In the numerical analyses presented in 
Fig.~\ref{softlpg.ab} and \ref{softlpg.nb}, we assume $\delta_{B-F} = 1$. 
We note that even if an additional suppresion 
$\delta_{B-F} \sim 0.1$ is present, with a  value of 
$Im(A) \simeq 10 \; TeV$ at the resonance our model can still account 
for the observed BAU.

\section{Conclusion}\label{concl}

We have shown in this paper that, in contrast to the predictions of models with 
lop-sided textures, the predictions for LFV decays are well below the current experimental 
bounds. This is demonstrated in a model based on SUSY SO(10) with symmetric mass 
textures which give rise to predictions for all fermion masses and mixing angles, 
including those in the neutrino sector, that are in good agreement with experimental data 
within $1 \sigma$. The predictions of our model for LFV processes, 
$\ell_{i} \rightarrow \ell_{j} \gamma$, $\mu-e$ conversion as well as 
$\mu \rightarrow 3e$, are well below the 
most stringent bounds up-to date. Our predictions for many processes 
are within the reach 
of the next generation of LFV searches. This is especially true 
for $\mu-e$ conversion and $\mu \rightarrow e \gamma$.
We have also investigated the possibility of baryogenesis resulting from 
soft leptogenesis. 
Our model predicts $M_{1} < 10^{9} GeV$ which is the
required condition for this mechanism to work.
With the soft SUSY masses assuming their natural 
values, $B^\prime \sim 1.4 \; TeV$ and $Im(A) \sim 1 \; TeV$, 
we find that our model can indeed accommodate the observed 
baryon asymmetry of the Universe.

\begin{acknowledgments}
M-CC and KTM are supported, in part, 
by the U.S. Department of Energy under Grant No. DE-AC02-98CH10886 and 
DE-FG03-95ER40892, respectively. M-CC would also like to acknowledge 
Aspen Center for Physics, where part of this work was done, for its 
hospitality and for providing a very stimulating atmosphere. 
\end{acknowledgments}

%\appendix

%\section{Vacuum alignment in the flavon sector}

\bibliography{lfv}

\begin{thebibliography}{55}
\expandafter\ifx\csname natexlab\endcsname\relax\def\natexlab#1{#1}\fi
\expandafter\ifx\csname bibnamefont\endcsname\relax
  \def\bibnamefont#1{#1}\fi
\expandafter\ifx\csname bibfnamefont\endcsname\relax
  \def\bibfnamefont#1{#1}\fi
\expandafter\ifx\csname citenamefont\endcsname\relax
  \def\citenamefont#1{#1}\fi
\expandafter\ifx\csname url\endcsname\relax
  \def\url#1{\texttt{#1}}\fi
\expandafter\ifx\csname urlprefix\endcsname\relax\def\urlprefix{URL }\fi
\providecommand{\bibinfo}[2]{#2}
\providecommand{\eprint}[2][]{\url{#2}}

\bibitem[{\citenamefont{Chen and
  Mahanthappa}(2003{\natexlab{a}})}]{Chen:2003zv}
\bibinfo{author}{\bibfnamefont{M.-C.} \bibnamefont{Chen}} \bibnamefont{and}
  \bibinfo{author}{\bibfnamefont{K.~T.} \bibnamefont{Mahanthappa}},
  \bibinfo{journal}{Int. J. Mod. Phys.} \textbf{\bibinfo{volume}{A18}},
  \bibinfo{pages}{5819} (\bibinfo{year}{2003}{\natexlab{a}}).

\bibitem[{\citenamefont{Chen and Mahanthappa}(2000)}]{Chen:2000fp}
\bibinfo{author}{\bibfnamefont{M.-C.} \bibnamefont{Chen}} \bibnamefont{and}
  \bibinfo{author}{\bibfnamefont{K.~T.} \bibnamefont{Mahanthappa}},
  \bibinfo{journal}{Phys. Rev.} \textbf{\bibinfo{volume}{D62}},
  \bibinfo{pages}{113007} (\bibinfo{year}{2000}); %.
%\bibitem[{\citenamefont{Chen and Mahanthappa}(2002)}]{Chen:2001pr}
%\bibinfo{author}{\bibfnamefont{M.-C.} \bibnamefont{Chen}} \bibnamefont{and}
%  \bibinfo{author}{\bibfnamefont{K.~T.} \bibnamefont{Mahanthappa}},
  \bibinfo{journal}{{\it ibid.}} \textbf{\bibinfo{volume}{D65}},
  \bibinfo{pages}{053010} (\bibinfo{year}{2002}); %.
%\bibitem[{\citenamefont{Chen and
%  Mahanthappa}(2003{\natexlab{b}})}]{Chen:2002pa}
%\bibinfo{author}{\bibfnamefont{M.-C.} \bibnamefont{Chen}} \bibnamefont{and}
%  \bibinfo{author}{\bibfnamefont{K.~T.} \bibnamefont{Mahanthappa}},
  \bibinfo{journal}{{\it ibid.}} \textbf{\bibinfo{volume}{D68}},
  \bibinfo{pages}{017301} (\bibinfo{year}{2003}{\natexlab{b}}).

\bibitem[{\citenamefont{Barbieri et~al.}(1997)\citenamefont{Barbieri, Hall,
  Raby, and Romanino}}]{Barbieri:1997ww}
\bibinfo{author}{\bibfnamefont{R.}~\bibnamefont{Barbieri}},
  \bibinfo{author}{\bibfnamefont{L.~J.} \bibnamefont{Hall}},
  \bibinfo{author}{\bibfnamefont{S.}~\bibnamefont{Raby}}, \bibnamefont{and}
  \bibinfo{author}{\bibfnamefont{A.}~\bibnamefont{Romanino}},
  \bibinfo{journal}{Nucl. Phys.} \textbf{\bibinfo{volume}{B493}},
  \bibinfo{pages}{3} (\bibinfo{year}{1997}).

\bibitem[{\citenamefont{Eidelman et~al.}(2004)}]{Eidelman:2004wy}
\bibinfo{author}{\bibfnamefont{S.}~\bibnamefont{Eidelman}} \bibnamefont{et~al.}
  (\bibinfo{collaboration}{Particle Data Group}), \bibinfo{journal}{Phys.
  Lett.} \textbf{\bibinfo{volume}{B592}}, \bibinfo{pages}{1}
  (\bibinfo{year}{2004}).

\bibitem[{\citenamefont{Fusaoka and Koide}(1998)}]{Fusaoka:1998vc}
\bibinfo{author}{\bibfnamefont{H.}~\bibnamefont{Fusaoka}} \bibnamefont{and}
  \bibinfo{author}{\bibfnamefont{Y.}~\bibnamefont{Koide}},
  \bibinfo{journal}{Phys. Rev.} \textbf{\bibinfo{volume}{D57}},
  \bibinfo{pages}{3986} (\bibinfo{year}{1998}); %.
%
%\bibitem[{\citenamefont{Hocker et~al.}(2001)\citenamefont{Hocker, Lacker,
%  Laplace, and Le~Diberder}}]{Hocker:2001xe}
\bibinfo{author}{\bibfnamefont{A.}~\bibnamefont{Hocker}},
  \bibinfo{author}{\bibfnamefont{H.}~\bibnamefont{Lacker}},
  \bibinfo{author}{\bibfnamefont{S.}~\bibnamefont{Laplace}}, \bibnamefont{and}
  \bibinfo{author}{\bibfnamefont{F.}~\bibnamefont{Le~Diberder}},
  \bibinfo{journal}{Eur. Phys. J.} \textbf{\bibinfo{volume}{C21}},
  \bibinfo{pages}{225} (\bibinfo{year}{2001}).

\bibitem[{\citenamefont{Charles et~al.}(2004)}]{Charles:2004jd}
\bibinfo{author}{\bibfnamefont{J.}~\bibnamefont{Charles}} \bibnamefont{et~al.}
  (\bibinfo{collaboration}{CKMfitter Group}), %(\bibinfo{year}{2004}),
  \eprint{hep-ph/0406184}; %.
%
%\bibitem[{\citenamefont{Bona et~al.}(2004)}]{Bona:2004sj}
\bibinfo{author}{\bibfnamefont{M.}~\bibnamefont{Bona}} \bibnamefont{et~al.}
  (\bibinfo{collaboration}{UTfit}), %(\bibinfo{year}{2004}),
  \eprint{hep-ph/0408079}.

\bibitem[{\citenamefont{Hall et~al.}(1994)\citenamefont{Hall, Rattazzi, and
  Sarid}}]{Hall:1993gn}
\bibinfo{author}{\bibfnamefont{L.~J.} \bibnamefont{Hall}},
  \bibinfo{author}{\bibfnamefont{R.}~\bibnamefont{Rattazzi}}, \bibnamefont{and}
  \bibinfo{author}{\bibfnamefont{U.}~\bibnamefont{Sarid}},
  \bibinfo{journal}{Phys. Rev.} \textbf{\bibinfo{volume}{D50}},
  \bibinfo{pages}{7048} (\bibinfo{year}{1994}).

\bibitem[{\citenamefont{Maltoni et~al.}(2004)\citenamefont{Maltoni, Schwetz,
  Tortola, and Valle}}]{Maltoni:2004ei}
\bibinfo{author}{\bibfnamefont{M.}~\bibnamefont{Maltoni}},
  \bibinfo{author}{\bibfnamefont{T.}~\bibnamefont{Schwetz}},
  \bibinfo{author}{\bibfnamefont{M.~A.} \bibnamefont{Tortola}},
  \bibnamefont{and} \bibinfo{author}{\bibfnamefont{J.~W.~F.}
  \bibnamefont{Valle}}, 
  \bibinfo{journal}{New J. Phys.} \textbf{\bibinfo{volume}{6}},
  \bibinfo{pages}{122} (\bibinfo{year}{2004}),
  \eprint{hep-ph/0405172}.

\bibitem[{\citenamefont{Bahcall et~al.}(2004)\citenamefont{Bahcall,
  Gonzalez-Garcia, and Pena-Garay}}]{Bahcall:2004ut}
\bibinfo{author}{\bibfnamefont{J.~N.} \bibnamefont{Bahcall}},
  \bibinfo{author}{\bibfnamefont{M.~C.} \bibnamefont{Gonzalez-Garcia}},
  \bibnamefont{and}
  \bibinfo{author}{\bibfnamefont{C.}~\bibnamefont{Pena-Garay}},
 \eprint{hep-ph/0406294}.

\bibitem[{\citenamefont{Diwan et~al.}(2003)}]{Diwan:2003bp}
\bibinfo{author}{\bibfnamefont{M.~V.} \bibnamefont{Diwan}}
  \bibnamefont{et~al.}, \bibinfo{journal}{Phys. Rev.}
  \textbf{\bibinfo{volume}{D68}}, \bibinfo{pages}{012002}
  (\bibinfo{year}{2003}).

\bibitem[{\citenamefont{Masiero et~al.}(2004)\citenamefont{Masiero, Vempati,
  and Vives}}]{Masiero:2004js}
For a recent review, see 
\bibinfo{author}{\bibfnamefont{A.}~\bibnamefont{Masiero}},
  \bibinfo{author}{\bibfnamefont{S.~K.} \bibnamefont{Vempati}},
  \bibnamefont{and} \bibinfo{author}{\bibfnamefont{O.}~\bibnamefont{Vives}}, 
   \eprint{hep-ph/0407325}.

\bibitem[{\citenamefont{Hisano et~al.}(1996)\citenamefont{Hisano, Moroi, Tobe,
  and Yamaguchi}}]{Hisano:1995cp}
\bibinfo{author}{\bibfnamefont{J.}~\bibnamefont{Hisano}},
  \bibinfo{author}{\bibfnamefont{T.}~\bibnamefont{Moroi}},
  \bibinfo{author}{\bibfnamefont{K.}~\bibnamefont{Tobe}}, \bibnamefont{and}
  \bibinfo{author}{\bibfnamefont{M.}~\bibnamefont{Yamaguchi}},
  \bibinfo{journal}{Phys. Rev.} \textbf{\bibinfo{volume}{D53}},
  \bibinfo{pages}{2442} (\bibinfo{year}{1996}); %.
%
%\bibitem[{\citenamefont{Hisano and Nomura}(1999)}]{Hisano:1998fj}
\bibinfo{author}{\bibfnamefont{J.}~\bibnamefont{Hisano}} \bibnamefont{and}
  \bibinfo{author}{\bibfnamefont{D.}~\bibnamefont{Nomura}},
  \bibinfo{journal}{Phys. Rev.} \textbf{\bibinfo{volume}{D59}},
  \bibinfo{pages}{116005} (\bibinfo{year}{1999}).

%\bibitem[{\citenamefont{Casas and Ibarra}(2001)}]{Casas:2001sr}
%\bibinfo{author}{\bibfnamefont{J.~A.} \bibnamefont{Casas}} \bibnamefont{and}
%  \bibinfo{author}{\bibfnamefont{A.}~\bibnamefont{Ibarra}},
%  \bibinfo{journal}{Nucl. Phys.} \textbf{\bibinfo{volume}{B618}},
%  \bibinfo{pages}{171} (\bibinfo{year}{2001}); %.
%
%\bibitem[{\citenamefont{Lavignac et~al.}(2001)\citenamefont{Lavignac, Masina,
%  and Savoy}}]{Lavignac:2001vp}
%\bibinfo{author}{\bibfnamefont{S.}~\bibnamefont{Lavignac}},
%  \bibinfo{author}{\bibfnamefont{I.}~\bibnamefont{Masina}}, \bibnamefont{and}
%  \bibinfo{author}{\bibfnamefont{C.~A.} \bibnamefont{Savoy}},
%  \bibinfo{journal}{Phys. Lett.} \textbf{\bibinfo{volume}{B520}},
%  \bibinfo{pages}{269} (\bibinfo{year}{2001}); %.
%
%\bibitem[{\citenamefont{Ellis et~al.}(2002)\citenamefont{Ellis, Hisano, Lola,
%  and Raidal}}]{Ellis:2001xt}
%\bibinfo{author}{\bibfnamefont{J.~R.} \bibnamefont{Ellis}},
%  \bibinfo{author}{\bibfnamefont{J.}~\bibnamefont{Hisano}},
%  \bibinfo{author}{\bibfnamefont{S.}~\bibnamefont{Lola}}, \bibnamefont{and}
%  \bibinfo{author}{\bibfnamefont{M.}~\bibnamefont{Raidal}},
%  \bibinfo{journal}{Nucl. Phys.} \textbf{\bibinfo{volume}{B621}},
%  \bibinfo{pages}{208} (\bibinfo{year}{2002}); %.
%
%\bibitem[{\citenamefont{Lavignac et~al.}(2002)\citenamefont{Lavignac, Masina,
%  and Savoy}}]{Lavignac:2002gf}
%\bibinfo{author}{\bibfnamefont{S.}~\bibnamefont{Lavignac}},
%  \bibinfo{author}{\bibfnamefont{I.}~\bibnamefont{Masina}}, \bibnamefont{and}
%  \bibinfo{author}{\bibfnamefont{C.~A.} \bibnamefont{Savoy}},
%  \bibinfo{journal}{Nucl. Phys.} \textbf{\bibinfo{volume}{B633}},
%  \bibinfo{pages}{139} (\bibinfo{year}{2002}); %.
%
%\bibitem[{\citenamefont{Ellis and Raidal}(2002)}]{Ellis:2002xg}
%\bibinfo{author}{\bibfnamefont{J.~R.} \bibnamefont{Ellis}} \bibnamefont{and}
%  \bibinfo{author}{\bibfnamefont{M.}~\bibnamefont{Raidal}},
%  \bibinfo{journal}{Nucl. Phys.} \textbf{\bibinfo{volume}{B643}},
%  \bibinfo{pages}{229} (\bibinfo{year}{2002}); %.
%
%\bibitem[{\citenamefont{Masiero et~al.}(2003)\citenamefont{Masiero, Vempati,
%  and Vives}}]{Masiero:2002jn}
%\bibinfo{author}{\bibfnamefont{A.}~\bibnamefont{Masiero}},
%  \bibinfo{author}{\bibfnamefont{S.~K.} \bibnamefont{Vempati}},
%  \bibnamefont{and} \bibinfo{author}{\bibfnamefont{O.}~\bibnamefont{Vives}},
%  \bibinfo{journal}{Nucl. Phys.} \textbf{\bibinfo{volume}{B649}},
%  \bibinfo{pages}{189} (\bibinfo{year}{2003}); %.
%
%\bibitem[{\citenamefont{Masiero et~al.}(2004)\citenamefont{Masiero, Vempati,
%  and Vives}}]{Masiero:2004js}
%\bibinfo{author}{\bibfnamefont{A.}~\bibnamefont{Masiero}},
%  \bibinfo{author}{\bibfnamefont{S.~K.} \bibnamefont{Vempati}},
%  \bibnamefont{and} \bibinfo{author}{\bibfnamefont{O.}~\bibnamefont{Vives}}, 
%   \eprint{hep-ph/0407325}.

\bibitem[{\citenamefont{Brooks et~al.}(1999)}]{Brooks:1999pu}
\bibinfo{author}{\bibfnamefont{M.~L.} \bibnamefont{Brooks}}
  \bibnamefont{et~al.} (\bibinfo{collaboration}{MEGA}), \bibinfo{journal}{Phys.
  Rev. Lett.} \textbf{\bibinfo{volume}{83}}, \bibinfo{pages}{1521}
  (\bibinfo{year}{1999}).

\bibitem[{\citenamefont{Barkov et~al.}(1999)}]{Barkov:1999}
\bibinfo{author}{\bibfnamefont{L.~M.} \bibnamefont{Barkov}}
  \bibnamefont{et~al.}, \bibinfo{journal}{Research Proposal to PSI,}
  \textbf{\bibinfo{volume}{http://www.icepp.s.u-tokyo.ac.jp/meg}}
  (\bibinfo{year}{1999}).

\bibitem[{\citenamefont{Bellgardt et~al.}(1988)}]{Bellgardt:1987du}
\bibinfo{author}{\bibfnamefont{U.}~\bibnamefont{Bellgardt}}
  \bibnamefont{et~al.} (\bibinfo{collaboration}{SINDRUM}),
  \bibinfo{journal}{Nucl. Phys.} \textbf{\bibinfo{volume}{B299}},
  \bibinfo{pages}{1} (\bibinfo{year}{1988}).

\bibitem[{\citenamefont{Wints}(1998)}]{Wints:1998}
\bibinfo{author}{\bibfnamefont{P.}~\bibnamefont{Wints}}
  (\bibinfo{collaboration}{SINDRUM II}), \bibinfo{journal}{in Proceedings of
  the First International Symposium on Lepton and Baryon Number Violation, ed.
  H. V. Klapdor-Kleingrothaus and I.V. Krivosheina,} p. \bibinfo{pages}{534}
  (\bibinfo{year}{1998}), \eprint{Institute of Physics Publishing, Bristol and
  Philadelphia}.

\bibitem[{\citenamefont{Bachman et~al.}(1997)}]{Bachman:1997}
\bibinfo{author}{\bibfnamefont{M.}~\bibnamefont{Bachman}} \bibnamefont{et~al.}
  (\bibinfo{collaboration}{MECO}), \bibinfo{journal}{Proposal to BNL,}
  \textbf{\bibinfo{volume}{http://meco.ps.uci.edu}} (\bibinfo{year}{1997}).

\bibitem[{\citenamefont{Inami}(2003)}]{Inami:2003}
\bibinfo{author}{\bibfnamefont{K.}~\bibnamefont{Inami}}
  (\bibinfo{collaboration}{BELLE}), \bibinfo{journal}{Talk presented at the
  19th International Workshop on Weak Interactions and Neutrinos (WIN'03), Lake
  Geneva, Wisconsin,}  (\bibinfo{year}{2003}).

\bibitem[{\citenamefont{Abe et~al.}(2004)}]{Abe:2003sx}
\bibinfo{author}{\bibfnamefont{K.}~\bibnamefont{Abe}} \bibnamefont{et~al.}
  (\bibinfo{collaboration}{Belle}), \bibinfo{journal}{Phys. Rev. Lett.}
  \textbf{\bibinfo{volume}{92}}, \bibinfo{pages}{171802}
  (\bibinfo{year}{2004}).

\bibitem[{\citenamefont{Petcov et~al.}(2004)\citenamefont{Petcov, Profumo,
  Takanishi, and Yaguna}}]{Petcov:2003zb}
\bibinfo{author}{\bibfnamefont{S.~T.} \bibnamefont{Petcov}},
  \bibinfo{author}{\bibfnamefont{S.}~\bibnamefont{Profumo}},
  \bibinfo{author}{\bibfnamefont{Y.}~\bibnamefont{Takanishi}},
  \bibnamefont{and} \bibinfo{author}{\bibfnamefont{C.~E.}
  \bibnamefont{Yaguna}}, \bibinfo{journal}{Nucl. Phys.}
  \textbf{\bibinfo{volume}{B676}}, \bibinfo{pages}{453} (\bibinfo{year}{2004}).

\bibitem[{\citenamefont{Bando et~al.}(2004)\citenamefont{Bando, Kaneko, Obara,
  and Tanimoto}}]{Bando:2004hi}
\bibinfo{author}{\bibfnamefont{M.}~\bibnamefont{Bando}},
  \bibinfo{author}{\bibfnamefont{S.}~\bibnamefont{Kaneko}},
  \bibinfo{author}{\bibfnamefont{M.}~\bibnamefont{Obara}}, \bibnamefont{and}
  \bibinfo{author}{\bibfnamefont{M.}~\bibnamefont{Tanimoto}}, 
  \eprint{hep-ph/0405071}.

\bibitem[{\citenamefont{Albright et~al.}(1998)\citenamefont{Albright, Babu, and
  Barr}}]{Albright:1998vf}
\bibinfo{author}{\bibfnamefont{C.~H.} \bibnamefont{Albright}},
  \bibinfo{author}{\bibfnamefont{K.~S.} \bibnamefont{Babu}}, \bibnamefont{and}
  \bibinfo{author}{\bibfnamefont{S.~M.} \bibnamefont{Barr}},
  \bibinfo{journal}{Phys. Rev. Lett.} \textbf{\bibinfo{volume}{81}},
  \bibinfo{pages}{1167} (\bibinfo{year}{1998}).

\bibitem[{\citenamefont{Barr}(2004)}]{Barr:2003fn}
\bibinfo{author}{\bibfnamefont{S.~M.} \bibnamefont{Barr}},
  \bibinfo{journal}{Phys. Lett.} \textbf{\bibinfo{volume}{B578}},
  \bibinfo{pages}{394} (\bibinfo{year}{2004}); %.
%
%\bibitem[{\citenamefont{Blazek and King}(2003)}]{Blazek:2002wq}
\bibinfo{author}{\bibfnamefont{T.}~\bibnamefont{Blazek}} \bibnamefont{and}
  \bibinfo{author}{\bibfnamefont{S.~F.} \bibnamefont{King}},
  \bibinfo{journal}{Nucl. Phys.} \textbf{\bibinfo{volume}{B662}},
  \bibinfo{pages}{359} (\bibinfo{year}{2003}).

\bibitem[{\citenamefont{Bennett et~al.}(2003)}]{Bennett:2003bz}
\bibinfo{author}{\bibfnamefont{C.~L.} \bibnamefont{Bennett}}
  \bibnamefont{et~al.}, \bibinfo{journal}{Astrophys. J. Suppl.}
  \textbf{\bibinfo{volume}{148}}, \bibinfo{pages}{1} (\bibinfo{year}{2003}).

\bibitem[{\citenamefont{Fukugita and Yanagida}(1986)}]{Fukugita:1986hr}
\bibinfo{author}{\bibfnamefont{M.}~\bibnamefont{Fukugita}} \bibnamefont{and}
  \bibinfo{author}{\bibfnamefont{T.}~\bibnamefont{Yanagida}},
  \bibinfo{journal}{Phys. Lett.} \textbf{\bibinfo{volume}{B174}},
  \bibinfo{pages}{45} (\bibinfo{year}{1986}); %.
%
%\bibitem[{\citenamefont{Luty}(1992)}]{Luty:1992un}
\bibinfo{author}{\bibfnamefont{M.~A.} \bibnamefont{Luty}},
  \bibinfo{journal}{Phys. Rev.} \textbf{\bibinfo{volume}{D45}},
  \bibinfo{pages}{455} (\bibinfo{year}{1992}); %.
%
%\bibitem[{\citenamefont{Buchmuller and Plumacher}(1996)}]{Buchmuller:1996pa}
\bibinfo{author}{\bibfnamefont{W.}~\bibnamefont{Buchmuller}} \bibnamefont{and}
  \bibinfo{author}{\bibfnamefont{M.}~\bibnamefont{Plumacher}},
  \bibinfo{journal}{Phys. Lett.} \textbf{\bibinfo{volume}{B389}},
  \bibinfo{pages}{73} (\bibinfo{year}{1996}).

\bibitem[{\citenamefont{Grossman et~al.}(2003)\citenamefont{Grossman, Kashti,
  Nir, and Roulet}}]{Grossman:2003jv}
\bibinfo{author}{\bibfnamefont{Y.}~\bibnamefont{Grossman}},
  \bibinfo{author}{\bibfnamefont{T.}~\bibnamefont{Kashti}},
  \bibinfo{author}{\bibfnamefont{Y.}~\bibnamefont{Nir}}, \bibnamefont{and}
  \bibinfo{author}{\bibfnamefont{E.}~\bibnamefont{Roulet}},
  \bibinfo{journal}{Phys. Rev. Lett.} \textbf{\bibinfo{volume}{91}},
  \bibinfo{pages}{251801} (\bibinfo{year}{2003}).

\bibitem[{\citenamefont{D'Ambrosio et~al.}(2003)\citenamefont{D'Ambrosio,
  Giudice, and Raidal}}]{D'Ambrosio:2003wy}
\bibinfo{author}{\bibfnamefont{G.}~\bibnamefont{D'Ambrosio}},
  \bibinfo{author}{\bibfnamefont{G.~F.} \bibnamefont{Giudice}},
  \bibnamefont{and} \bibinfo{author}{\bibfnamefont{M.}~\bibnamefont{Raidal}},
  \bibinfo{journal}{Phys. Lett.} \textbf{\bibinfo{volume}{B575}},
  \bibinfo{pages}{75} (\bibinfo{year}{2003}).


%\cite{Boubekeur:2002jn}
\bibitem{Boubekeur:2002jn}
L.~Boubekeur, 
%``Leptogenesis at low scale,''
hep-ph/0208003; 
%%CITATION = HEP-PH 0208003;%%
L.~Boubekeur, T.~Hambye and G.~Senjanovic,
%``Low-scale leptogenesis and soft supersymmetry breaking,''
Phys. Rev. Lett. 93, 111601 (2004),  
hep-ph/0404038.
%%CITATION = HEP-PH 0404038;%%

\bibitem[{\citenamefont{Buchmuller et~al.}(2002)\citenamefont{Buchmuller,
  Di~Bari, and Plumacher}}]{Buchmuller:2002jk}
\bibinfo{author}{\bibfnamefont{W.}~\bibnamefont{Buchmuller}},
  \bibinfo{author}{\bibfnamefont{P.}~\bibnamefont{Di~Bari}}, \bibnamefont{and}
  \bibinfo{author}{\bibfnamefont{M.}~\bibnamefont{Plumacher}},
  \bibinfo{journal}{Phys. Lett.} \textbf{\bibinfo{volume}{B547}},
  \bibinfo{pages}{128} (\bibinfo{year}{2002}); %.
%
%\bibitem[{\citenamefont{Giudice et~al.}(2004)\citenamefont{Giudice, Notari,
%  Raidal, Riotto, and Strumia}}]{Giudice:2003jh}
\bibinfo{author}{\bibfnamefont{G.~F.} \bibnamefont{Giudice}},
  \bibinfo{author}{\bibfnamefont{A.}~\bibnamefont{Notari}},
  \bibinfo{author}{\bibfnamefont{M.}~\bibnamefont{Raidal}},
  \bibinfo{author}{\bibfnamefont{A.}~\bibnamefont{Riotto}}, \bibnamefont{and}
  \bibinfo{author}{\bibfnamefont{A.}~\bibnamefont{Strumia}},
  \bibinfo{journal}{Nucl. Phys.} \textbf{\bibinfo{volume}{B685}},
  \bibinfo{pages}{89} (\bibinfo{year}{2004}).

\bibitem[{\citenamefont{Grossman et~al.}(2004)\citenamefont{Grossman, Kashti,
  Nir, and Roulet}}]{Grossman:2004dz}
\bibinfo{author}{\bibfnamefont{Y.}~\bibnamefont{Grossman}},
  \bibinfo{author}{\bibfnamefont{T.}~\bibnamefont{Kashti}},
  \bibinfo{author}{\bibfnamefont{Y.}~\bibnamefont{Nir}}, \bibnamefont{and}
  \bibinfo{author}{\bibfnamefont{E.}~\bibnamefont{Roulet}}
  (\bibinfo{year}{2004}), \eprint{hep-ph/0407063}.


\bibitem[{\citenamefont{Ross and Sarkar}(1996)}]{Ross:1995dq}
M.~Y.~Khlopov and A.~D.~Linde,
%``Is It Easy To Save The Gravitino?,''
Phys.\ Lett.\ B {\bf 138}, 265 (1984); 
%%CITATION = PHLTA,B138,265;%%
\bibinfo{author}{\bibfnamefont{B.~A.} \bibnamefont{Campbell}},
  \bibinfo{author}{\bibfnamefont{S.}~\bibnamefont{Davidson}}, \bibnamefont{and}
  \bibinfo{author}{\bibfnamefont{K.~A.} \bibnamefont{Olive}},
  \bibinfo{journal}{Nucl. Phys.} \textbf{\bibinfo{volume}{B399}},
  \bibinfo{pages}{111} (\bibinfo{year}{1993}); 
\bibinfo{author}{\bibfnamefont{S.}~\bibnamefont{Sarkar}}, 
  \eprint{hep-ph/9510369}; %.
\bibinfo{author}{\bibfnamefont{G.~G.} \bibnamefont{Ross}} \bibnamefont{and}
  \bibinfo{author}{\bibfnamefont{S.}~\bibnamefont{Sarkar}},
  \bibinfo{journal}{Nucl. Phys.} \textbf{\bibinfo{volume}{B461}},
  \bibinfo{pages}{597} (\bibinfo{year}{1996}); 
M.~Y.~Khlopov,
{\it Cosmoparticle physics}, 577p, World Scientific, Singapore (1999).

\bibitem[{\citenamefont{Covi et~al.}(1998)\citenamefont{Covi, Rius, Roulet, and
  Vissani}}]{Covi:1997dr}
\bibinfo{author}{\bibfnamefont{L.}~\bibnamefont{Covi}},
  \bibinfo{author}{\bibfnamefont{N.}~\bibnamefont{Rius}},
  \bibinfo{author}{\bibfnamefont{E.}~\bibnamefont{Roulet}}, \bibnamefont{and}
  \bibinfo{author}{\bibfnamefont{F.}~\bibnamefont{Vissani}},
  \bibinfo{journal}{Phys. Rev.} \textbf{\bibinfo{volume}{D57}},
  \bibinfo{pages}{93} (\bibinfo{year}{1998}).

\bibitem[{\citenamefont{Kolb and Turner}(1990)}]{Kolb:1990vq}
\bibinfo{author}{\bibfnamefont{E.~W.} \bibnamefont{Kolb}} \bibnamefont{and}
  \bibinfo{author}{\bibfnamefont{M.~S.} \bibnamefont{Turner}},
  \bibinfo{journal}{The Early Universe,}  (\bibinfo{year}{1990}),
  \bibinfo{note}{Redwood City, USA: Addison-Wesley 547 p. (Frontiers in
  physics, 69)}; %.
%
%\bibitem[{\citenamefont{Nielsen and Takanishi}(2001)}]{Nielsen:2001fy}
\bibinfo{author}{\bibfnamefont{H.~B.} \bibnamefont{Nielsen}} \bibnamefont{and}
  \bibinfo{author}{\bibfnamefont{Y.}~\bibnamefont{Takanishi}},
  \bibinfo{journal}{Phys. Lett.} \textbf{\bibinfo{volume}{B507}},
  \bibinfo{pages}{241} (\bibinfo{year}{2001}).

\bibitem[{\citenamefont{Yamaguchi and Yoshioka}(2002)}]{Yamaguchi:2002zy}
\bibinfo{author}{\bibfnamefont{M.}~\bibnamefont{Yamaguchi}} \bibnamefont{and}
  \bibinfo{author}{\bibfnamefont{K.}~\bibnamefont{Yoshioka}},
  \bibinfo{journal}{Phys. Lett.} \textbf{\bibinfo{volume}{B543}},
  \bibinfo{pages}{189} (\bibinfo{year}{2002}).

\end{thebibliography}

\end{document}